\documentclass[prd,reprint,
nofootinbib,notitlepage,aps,tightenlines,
preprintnumbers,amsmath,amssymb,amsfonts,showpacs,superscriptaddress]{revtex4-2}

\usepackage[hyperindex,breaklinks]{hyperref}
\usepackage{graphicx,epstopdf}
\usepackage{comment}
\usepackage{caption,subcaption}
\usepackage{mathtools}
\usepackage{amsmath}
\usepackage{booktabs}
\usepackage{dcolumn}% Align table columns on decimal point
\usepackage{natbib,bm}
\usepackage[usenames,dvipsnames]{color}
\usepackage{slashed}    
\usepackage[noabbrev]{cleveref}
\usepackage{multirow}
\usepackage{chngcntr}

\def\be{\begin{equation}}
\def\ee{\end{equation}}
\def\ba{\begin{eqnarray}}
\def\ea{\end{eqnarray}}
\def\ge{\mathrel{\raise.3ex\hbox{$>$\kern-.75em\lower1ex\hbox{$\sim$}}}}
\def\la{\mathrel{\raise.3ex\hbox{$<$\kern-.75em\lower1ex\hbox{$\sim$}}}}

\def\thesection{\arabic{section}}
\def\theequation{\arabic{equation}}
\def\simgt{\mathrel{\raise.3ex\hbox{$>$\kern-.75em\lower1ex\hbox{$\sim$}}}}
\def\simlt{\mathrel{\raise.3ex\hbox{$<$\kern-.75em\lower1ex\hbox{$\sim$}}}}
\newcommand{\No}{N_{\mathrm{opt}}}
\newcommand{\lmn}{{\ell m n}}

\newcommand{\nc}{\newcommand}

\nc{\gone}{\bar g_{\pi NN}^{(1)}}
\nc{\gzero}{\bar g_{\pi NN}^{(0)}}
\nc{\al}{\alpha}
\nc{\ga}{\gamma}
\nc{\de}{\delta}
\nc{\ep}{\epsilon}
\nc{\ze}{\zeta}
\nc{\et}{\eta}
\nc{\ka}{\kappa}
\nc{\rh}{\rho}
\nc{\si}{\sigma}
\nc{\ta}{\tau}
\nc{\up}{\upsilon}
\nc{\ph}{\phi}
\nc{\ch}{\chi}
\nc{\ps}{\psi}
\nc{\om}{\omega}
\nc{\Ga}{\Gamma}
\nc{\De}{\Delta}
\nc{\La}{\Lambda}
\nc{\Si}{\Sigma}
\nc{\Up}{\Upsilon}
\nc{\Ph}{\Phi}
\nc{\Ps}{\Psi}
\nc{\Om}{\Omega}
\nc{\ptl}{\partial}
\nc{\del}{\nabla}
\nc{\ov}{\overline}
\nc{\newcaption}[1]{\centerline{\parbox{15cm}{\caption{#1}}}}
\nc{\us}{U(1)$_S$}

\def\beq{\begin{equation}}
\def\eeq{\end{equation}}
\def\bmat{\begin{displaymath}}
\def\emat{\end{displaymath}}
\def\bear{\begin{eqnarray}}
\def\eear{\end{eqnarray}}
\def\ba{\begin{eqnarray}}
\def\ea{\end{eqnarray}}
\def\bery{\begin{array}}
\def\ery{\end{array}}
\def\bit{\begin{itemize}}
\def\eit{\end{itemize}}
\def\ben{\begin{enumerate}}
\def\een{\end{enumerate}}
\def\btab{\begin{tabular}}
\def\etab{\end{tabular}}
\def\btbl{\begin{table}}
\def\etbl{\end{table}}
\def\bfig{\begin{figure}[htb]}
\def\efig{\end{figure}}
\def\bpic{\begin{picture}}
\def\epic{\end{picture}}

\def\nnl{\nonumber \\}

\def\ga{\mathrel{\raise.3ex\hbox{$>$\kern-.75em\lower1ex\hbox{$\sim$}}}}
\def\la{\mathrel{\raise.3ex\hbox{$<$\kern-.75em\lower1ex\hbox{$\sim$}}}}
\def\gappeq{\mathrel{\rlap {\raise.5ex\hbox{$>$}}
{\lower.5ex\hbox{$\sim$}}}}
\def\lappeq{\mathrel{\rlap{\raise.5ex\hbox{$<$}}
{\lower.5ex\hbox{$\sim$}}}}

\def\gyr{{\rm \, G\kern-0.125em yr}}
\def\mev{{\rm \, Me\kern-0.125em V}}
\def\gev{{\rm \, Ge\kern-0.125em V}}
\def\tev{{\rm \, Te\kern-0.125em V}}

\begin{document}

\title{Ringdown signatures in the Ernst-Wild geometry: modeling Kerr black holes immersed in a magnetic field}%

\author{Kate J. Taylor}
\affiliation{Department of Physics and Astronomy, University of Victoria,
Victoria, BC V8P 5C2, Canada}
\author{Adam Ritz}
%\thanks{aritz@uvic.ca, corresponding author}
\affiliation{Department of Physics and Astronomy, University of Victoria,
Victoria, BC V8P 5C2, Canada}

\date{June 12, 2024}

\begin{abstract}
\noindent
We analyze the quasinormal mode spectrum for Kerr black holes surrounded by an asymptotically uniform magnetic field, modeled with the Ernst-Wild geometry. A perturbative expansion in both the rotation parameter $a$ and the magnetic field $B$ allows the analysis of perturbations with Kerr-like asymptotics well inside the Melvin radius, and we obtain the spectrum for a variety of scalar quasinormal modes over a range of parameters using the continued fraction method. We then interpolate the low-lying mode spectrum to construct an Ernst-Wild template for the ringdown, and use the LIGO-Virgo-KAGRA analysis tool \texttt{pyRing} to assess the impact of the magnetosphere on the extraction of ringdown signatures from several observed binary black hole mergers. 
\end{abstract}

\maketitle

%%%%%%%%%%%%%%%%%%%%%%%%%%%%%%%%%%%%%%%%%%%%%%%%%%%%%%%%%%%%%%%
\section{Introduction}\label{sec:Introduction}
%%%%%%%%%%%%%%%%%%%%%%%%%%%%%%%%%%%%%%%%%%%%%%%%%%%%%%%%%%%%%%%

The characteristic exponential ringdown signature of the black hole remnant of a binary merger, as predicted by General Relativity, is readily apparent in the first LIGO-Virgo-KAGRA (LVK) event GW150914 \cite{LIGOScientific:2016lio}. The signature closely matches the expectations of the complex quasinormal mode (QNM) pseudospectrum following from a linearized analysis of Kerr black hole perturbations (for reviews, see \cite{Kokkotas:1999bd,Berti:2009kk,Konoplya:2011qq}). While the fully dynamical post-merger behaviour is in many ways more complex than the naive setting for linear perturbation theory about a stationary black hole configuration, the data clearly indicates consistency to leading order with an underlying picture of dissipative ringdown, as anticipated due to the presence of a horizon. With the growing catalogue of binary merger events allowing more detailed use of the ringdown component to extract black hole parameters, the extent to which individual (linear) quasinormal modes can be isolated is coming under increasing scrutiny \cite{CalderonBustillo:2020rmh,Capano:2021etf,Cotesta:2022pci,Isi:2022mhy,Finch:2022ynt,Cheung:2022rbm,Isi:2023nif,Wang:2023xsy,Carullo:2023gtf,Zhu:2023mzv,Baibhav:2023clw,Nee:2023osy,Correia:2023bfn,Qiu:2023lwo,Zhu:2024dyl,may2024nonlinear}.
Presently, it is unclear if non-linearities arising from contamination of the merger itself could affect the experimental extraction of the waveform parameters. It is thought that these effects should not impact the extraction of the dominant $(2,2, 0)$ QNM, but it is not clear how these effects could impact the reliable measurements of higher overtones \cite{Zhu:2023mzv,cardoso2024physical}. Care is needed in adding higher-order modes, given the potential to `overfit' the data, which could lead to inaccurate values for the ringdown parameters.

A distinct, but related, question is how the basic Kerr template for these modes could be impacted theoretically by the black hole environment.
Early examples include scalar hair \cite{Bekenstein:1973mi, Bekenstein:1996pn}, and black hole charge within the full Kerr-Newman geometry \cite{PhysRevD.88.064048, Konoplya:2013rxa}; see also \cite{Barausse:2014tra,cannizzaro2024impact,Rahman:2022fay}. While these generalizations were considered primarily as a way to test for any deviation from the Kerr spectrum, in this paper we will explore the impact of a more physical black hole environment, namely an asymptotically uniform magnetosphere as an approximation to the field structure sustained by an accretion disk. While an astrophysically relevant magnetosphere is not expected to produce large corrections to the Kerr QNM spectrum, and thus seems impractical to detect for black holes in the LVK window given the current level of precision \cite{gupta2024possible}, it may still provide a nontrivial `nuisance' parameter that could impact the extraction of mass and spin within the Kerr waveform template. In particular, the QNM spectrum is known to be highly (and nonlinearly) sensitive to perturbations of the underlying potential, as characterized by the pseudospectrum for example \cite{Nollert_1996,Jaramillo_2021,cardoso2024physical}. Accordingly, our main goal will be to explore the impact of a magnetosphere on the modeling of ringdown from a merger via quasinormal modes.

Black hole magnetospheres have been analyzed extensively, with perhaps the most notable motivation being their potential role as power sources for active galaxies, e.g.\ via the Blandford-Znajek (BZ) mechanism \cite{Blandford:1977ds}. However, such energy extraction phenomena rely on the specifics of force-free electrodynamics in the black hole ergosphere, for which no analytic solutions are known beyond the perturbative regime of slow rotation. Thus, in considering ringdown, it is of interest to study cases where exact solutions do exist, and where the back-reaction on the black hole geometry can be fully taken into account. While the magnetized Kerr-Newman solution is technically the simplest, a more physical system requires an approximately uniform field away from the black hole. The regime where the energy density in the magnetic field is dominant is described by the Melvin solution \cite{Melvin:1965zza}, in which the spacetime asymptotics are also modified, which in general will completely alter the behaviour of the QNM spectrum. A Kerr black hole immersed in an asymptotically uniform magnetic field is described by the Ernst-Wild solution \cite{ErnstWild:1976Metric,Aliev:1989wz}, generalizing the static Ernst solution \cite{Ernst:1976mzr}. The QNM spectrum for this system was studied in \cite{Brito:2014super} focusing on the regime of very large magnetic fields, where the Melvin-like asymptotics at large radius lead to the possibility of superradiance \cite{Dolan_2013,Brito:2015oca,Brito_2020}. Our focus will instead be on the regime of small magnetic fields (relative to the black hole mass), which is more readily realized astrophysically. 

To make contact with astrophysical systems, where the magnetosphere is localized and the spacetime is asymptotically flat, we choose to expand perturbatively in the magnetic field strength $B$. On one hand, this is consistent with the small scale of observed astrophysical magnetic fields (in Planck units), but it also creates an important scale separation in the geometry. The non-asymptotically flat Melvin spacetime only becomes dominant outside the Melvin radius $r_M \sim 1/B \gg r_+$. Thus, for parametrically small $B$, the mode solutions for perturbations exhibit conventional Kerr-like asymptotics within a large intermediate interval in radius $r_+ \ll r \ll r_M$. More technically, the equations for perturbations of the spacetime exhibit a transition from Kerr-like asymptotics at ${\cal O}(B^2)$ to Melvin-like confining asymptotics at ${\cal O}(B^4)$. As a result, at ${\cal O}(B^2)$ the modes are primarily sensitive to the localized impact of the magnetic field near the Kerr horizon and photon sphere. Analyzing the spectrum of perturbations at this order also has the secondary benefit of technical simplicity. The analysis of linearized perturbations in the Ernst-Wild geometry is complicated by the fact that there is no known decoupling of the mode equations, which therefore need to be analyzed via complex numerical techniques \cite{Konoplya:2011qq,Pani:2013pma,Brenner_2021,Ghosh:2023etd} even in the scalar case. 
However, working at ${\cal O}(B^2)$ leads to simplifications, which we exploit in this  exploratory study by also working perturbatively in the Kerr rotation parameter $a$. This allows for complete decoupling of the perturbation equations.

We proceed to determine the scalar Ernst-Wild QNM spectrum as a function of the rotation parameter $a$ and the magnetic field $B^2$ for the low-lying modes most relevant for the ringdown signature. 
For simplicity, we choose to focus on scalar perturbations which, while unphysical, are known in the Kerr limit to behave similarly to the tensor modes. We quantify the impact of this assumption and the perturbative expansion in $B^2$ and $a$ in the next section, focusing on those modes of most interest in an application to mergers in the LVK catalogue. To gain some qualitative insight into the role the magnetic field strength $B$ in post-merger ringdown, we imagine excising the exterior region of the geometry beyond the Melvin radius and construct a scalar analogue of an Ernst-Wild ringdown template by extrapolating the QNM spectrum to larger values of $B$.
Utilizing the public LVK ringdown pipeline \texttt{pyRing}, we assess its fit to a number of merger events in the catalogue. More precisely, we vary the magnetic field and explore its impact on extracting the primary parameters of mass and spin for the final black hole remnant. We conclude that while the effect is not large, it warrants consideration in a similar manner to the potential impact of nonlinear corrections to the ringdown.

The rest of this paper is structured as follows. Section~2 reviews the Ernst-Wild geometry for the Kerr black hole immersed in an asymptotically uniform magnetic field, and the equation for scalar quasinormal mode perturbations. Further details of the large radius asymptotics within the perturbative expansion in $B$ are presented in Appendix A. Section~3 determines several quasinormal mode frequencies to leading order in the rotation and magnetic field parameters, and provides an interpolation that is then used to extrapolate over an extended range of remnant parameters. Section~4 describes the ensuing construction of an EW template, its implementation within the \texttt{pyRing} pipeline, and explores how a magnetosphere impacts the extraction of remnant mass and spin from several binary merger events in the LVK catalogue. The results are briefly discussed in Section~5, along with an outlook for future improvements in the analysis.

Planck units with $G=c=1$ are used throughout this work.

%%%%%%%%%%%%%%%%%%%%%%%%%%%%%%%%%%%%%%%%%%%%%%%%%%%%%%%%%%%%%%%%%%
\section{Perturbations of the Ernst-Wild solution}
\label{sec:Perturbations_of_EW}
%%%%%%%%%%%%%%%%%%%%%%%%%%%%%%%%%%%%%%%%%%%%%%%%%%%%%%%%%%%%%%%%%%%
Rotating black holes solutions immersed in asymptotically uniform magnetic fields are astrophysically well-motivated but necessarily complex, and it is remarkable that an analytic solution exists. The solution for a fully back-reacted Kerr black hole immersed in a magnetized Melvin geometry was initially obtained by Ernst and Wild \cite{ErnstWild:1976Metric}, following the use of a Harrison transformation \cite{Harrison:1968HarrisonTransformations} on a seed Kerr-Newman metric with mass $M$, angular momentum $J$ (with rotation parameter $a\equiv J/M$), and charge $q$. The initial solution was subsequently corrected and analyzed in some detail \cite{Gibbons:2013EW,Diaz:1985EW,Hedja:2015BHMagnetic,Astorino_2016,Brenner_2021}.

One of the intriguing features of the Ernst-Wild solution is that for the full geometry to be electrically neutral, the black hole necessarily acquires an induced Kerr-Newman electric charge $q=-2aM B$ via the Wald mechanism \cite{Wald:1974np}. 
Following \cite{Brito:2014super}, to second order in the spin parameter $ a$, and taking $q = -2 a M B $, the Ernst-Wild geometry takes the form \cite{Gibbons:2013EW,ErnstWild:1976Metric,Diaz:1985EW,Hedja:2015BHMagnetic},
\begin{align}
    \label{e:EW_metric}
	ds^2|_{{\tilde a}^2} = \varrho ^2\, \Lambda \biggr [ -\frac{\Delta}{\Sigma}dt^2 &+ \left ( \frac{dr^2}{\Delta} + d \theta^2 \right ) \biggr ] \nonumber \\ 
    &+ \frac{\Sigma\, \sin^2 \theta}{\varrho ^2\, \Lambda} (\Lambda_0 d\phi - \varpi dt )^2,
\end{align}
where $\Lambda_0$ is used to remove the conical singularity at the poles \cite{Hiscock:1981EWFixed}, and 
\begin{subequations}
    \label{e:EW_geometryterms}
        \begin{eqnarray}
        \label{e:EW_delta}
	   \Delta &=& r^2 - 2Mr + a^2 + 4 a^2 M^2 B^2 , \\
        \label{e:EW_sigma}
	   \varrho^2 &=& r^2 +  a ^2 \cos^2 \theta, \\
        \label{e:EW_tildeA}
	   \Sigma &=& r^4 + r a^2  \left ( (2M -r) \sin^2 \theta + 2r \right ), \\
        \nonumber
	   \Lambda &=& 1 + \frac{1}{2}B^2 r^2 \sin^2 \theta + \frac{1}{16} B^4 r^4 \sin^4 \theta\\ \nonumber
        &+&  a^2 \left [ \frac{1}{8} B^6 M^2 r^2\sin^2 2\theta \right . + \frac{1}{8} B^4 \{ 2Mr \sin^6 \theta \\ \nonumber
        &+& 2M \cos^2 \theta (M \cos^4 \theta + 2 (M - 2r)  \cos^2 \theta+ 9M + 8r)\\ \nonumber
	   &-&  8Mr + r^2 \sin^4 \theta \} \\
        \label{e:EW_H}
        &+& \left . \frac{B^2}{2 r}(r - M (7 + \cos 2 \theta)) \sin^2 \theta  \right ], \\
         \label{e:EW_H0}
	   \Lambda_0& \equiv& \Lambda(r, \theta = 0) = 1 + 3  a^2 M^2 B^4, \\
	   \varpi &=& \frac{ a M}{64 r^3} \left [ - B^4 r^3 (12 \cos 2 \theta + \cos 4 \theta ) (r - 2M) \right . \nonumber \\
        &+& \left . B^2 r^2 (256 - B^2 r (186 M + 51r )) + 128\right ].
    \end{eqnarray}
\end{subequations}

For our purposes, it will be useful to introduce a rescaled rotation parameter,
\begin{align}
 \tilde a \equiv  \frac{a}{M}= \frac{J}{M^2},
 \end{align}
 and a rescaled magnetic field,
\begin{align}
\tilde{B} \equiv BM,
\end{align}
where $B$ is the magnitude of the asymptotically uniform magnetic field strength. We will assume that the magnetic field surrounding the black hole is `weak', namely that $B M \ll 1$, a condition that is readily satisfied by known astrophysical magnetospheres. As discussed above, this restriction also has the benefit of restoring Kerr-like asymptotics for perturbations when we formally truncate the geometry at ${\cal O}(\tilde{B}^2)$, as illustrated in Appendix A. On restoring units, this limit takes the form \cite{piotrovich2010magnetic,Frolov:2010mi,Frolov_2012},
 \begin{align}
    \label{e:weakly_magntized}
    B \ll B_{M} = \frac{c^4}{G^{3/2} M_\odot} \left ( \frac{M_\odot}{M} \right ) \sim 10^{19}\left ( \frac{M_\odot}{M} \right ) {\rm Gauss} .
\end{align}

With these conventions, we can write the horizon radius $r_+$ and angular velocity $\Omega_H{=} - g_{t \phi}/g_{\phi \phi}|_{r=r_+}$ to $\mathcal O (\tilde a^2, \tilde B^2)$ in the form,
\begin{align}
    r_+ &= 2 M- \tilde a^2 \left(\frac{M}{2} +2 \tilde B^2 M \right), \\
    \Omega_H &=  \left .\frac{2 \tilde a M^2+4 \tilde a  \tilde B^2 r^2}{r^3}  \right |_{r=r_+}.
\end{align}
Note that the Kerr metric can be obtained by setting $B=0$, the Ernst metric can be obtained by setting $a =0$, and the Schwarzschild metric follows on setting $B=0$ and $a=0$.

%%%%%%%%%%%%%%%%%%%%%%%%%%%%%%%%%%%%%%%%%%%%%
\subsection{Linearized perturbation analysis} 
%%%%%%%%%%%%%%%%%%%%%%%%%%%%%%%%%%%%%%%%%%%%%
\begin{figure*}[tbp]
  \centering
  \includegraphics[width=\linewidth]{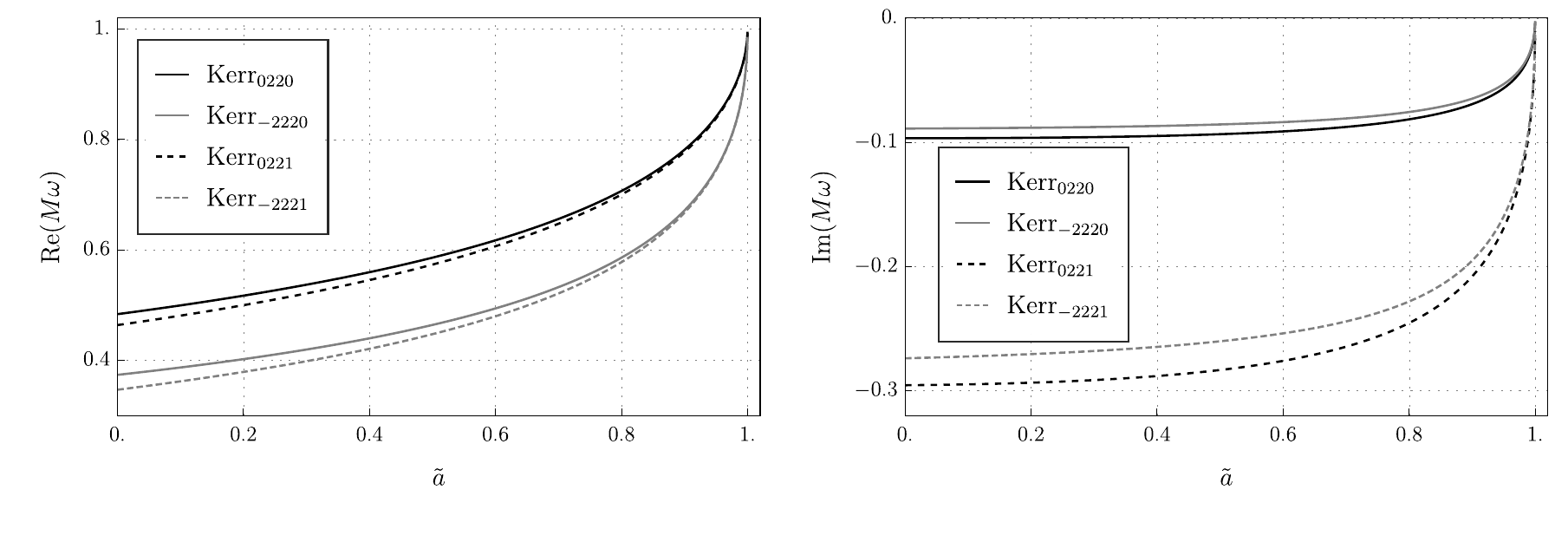}  
\vspace*{-0.5cm}
\caption{The real (left) and imaginary (right) parts of the leading scalar $s=0$ and tensor $s=-2$ Kerr quasinormal mode frequencies $\omega_{s \lmn}$ are shown as a function of the rotation parameter $\tilde{a}$ computed via the continued fraction technique.}
\label{fig:Kerr_scalar_tensor}
\end{figure*}
We now proceed to an analysis of linearized perturbations of the EW geometry. Our focus will be on the quasinormal mode spectrum, for use in modelling the ringdown phase of a magnetized rotating black hole.
In this exploratory study, we will make two important simplifying approximations. The first is to restrict attention to scalar rather than tensor perturbations.
\begin{enumerate}
    \item[(A1)] {\it Scalar rather than tensor perturbations}: This choice is unphysical for analyzing gravitational ringdown, but simplifies the analysis considerably, and for the dominant mode and subdominant overtone with mode numbers $(\ell,m,n)=(2,2,0)$ and $(2,2,1)$ respectively, the scalar and tensor quasinormal mode frequencies for the Kerr solution are remarkably similar quantitatively. This is  illustrated in Fig.~\ref{fig:Kerr_scalar_tensor}.
\end{enumerate}
This assumption allows us to focus on the Klein-Gordon equation for a massless scalar field,
\begin{align}
	\label{e:KG_scalar}
    \Box \Phi \equiv \frac{1}{\sqrt{-g}} \ptl_\mu (\sqrt{-g}g^{\mu \nu} \ptl_\nu \Phi) =0.
\end{align}
Following \cite{Dolan_2013,Brito:2014super}, for geometries such as EW, the radial and angular parts of the equation can be separated at the expense of introducing couplings between the different multipoles $\ell$. We use the following ansatz
\begin{align}
    \label{e:ansatzScalarField}
    \Phi(t, r, \theta, \phi) = \sum_{j, m} \frac{\bar{R}_{jm}(r)}{r} e^{-i\om t}\, Y_{j m }(\theta, \phi), 
\end{align}
where the $Y_{jm}$ are the usual spherical harmonics. The axisymmetry of the geometry leads to a degeneracy in the spectrum as a function of the azimuthal quantum number $m$. 

Substitution of this ansatz into Eq.~\ref{e:KG_scalar}, and using the orthogonality of the spherical harmonics, allows the reduction to a system of nine coupled radial equations \cite{Brito:2014super}. As noted above, we further assume that the magnetic field surrounding the black hole is weak,  $\tilde{B} = B M \ll 1$, and truncating the system of coupled radial equations at ${\cal O}(\tilde{a}^2,\tilde{B}^2)$, we have
\begin{equation}
 \frac{d^2 \bar{R}_{\ell m}}{d r_*^2} + \sum_{i=-1}^{1} V_{\ell+2i,m} \bar{R}_{\ell+2i,m} =0,
    \label{e:decoupled}
\end{equation}
where the tortoise coordinate is defined by $dr/dr^* = \De/(r^2 + a^2)$, and the potentials are given by,
\begin{equation}
    \begin{split}
    \mathllap{V_{\ell} } &= \omega^2 - \frac{dr}{dr_*} \biggr \{\frac{\ell (\ell + 1)}{r^2} + \frac{2M }{r^3} + \frac{\tilde B^2 m^2}{M^2} \biggr \}  \\
      &\qquad  - \tilde a m \omega  \left ( \frac{4 M^2}{r^3} + \frac{8 \tilde  B^2}{r} \right )    \\
      &\qquad + \tilde a^2 \left ( \mathcal V_0 + \mathcal V_2\, c_{\ell, \ell, m}^{(2)}  \right ),\\
    \mathllap{V_{\ell \pm 2}} &=  \tilde a^2 \left ( \mathcal V_2\, c_{\ell\pm 2, \ell, m}^{(2)}\right ), 
    \end{split}
\end{equation}
where
\begin{align}
	\nonumber
	\mathcal V_0 &= \frac{\tilde B^2 ( 8M^2-4 \ell (\ell + 1) M^2 + m^2r(r+4M))}{r^4} \nnl
    &- \frac{24 \tilde B^2 M^3}{r^5} - \frac{24M^4}{r^6} + \frac{M^2}{r^5} \biggr [\ell (\ell + 1)(r-4M) \nnl
    &+ r(m^2-(r - 2M)r \omega^2 - 1) + 12M \biggr ], \\
	\mathcal V_2 &= - \frac{\tilde B^2 m^2 (r- 2M) }{r^3} + \frac{(r - 2M) M^2 \omega^2 }{r^3}.
\end{align}
The Clebsch-Gordon coefficients, $c^{(2)}_{j, \ell, m}{\equiv}\langle \ell m | \cos^{2}\theta | j m \rangle$, vanish
unless $j = \ell -2,\, \ell,\, \ell + 2 $.

\begin{figure*}[tbp]
\begin{subfigure}{\textwidth}
  \centering
  % include first image
  \includegraphics[width=\linewidth]{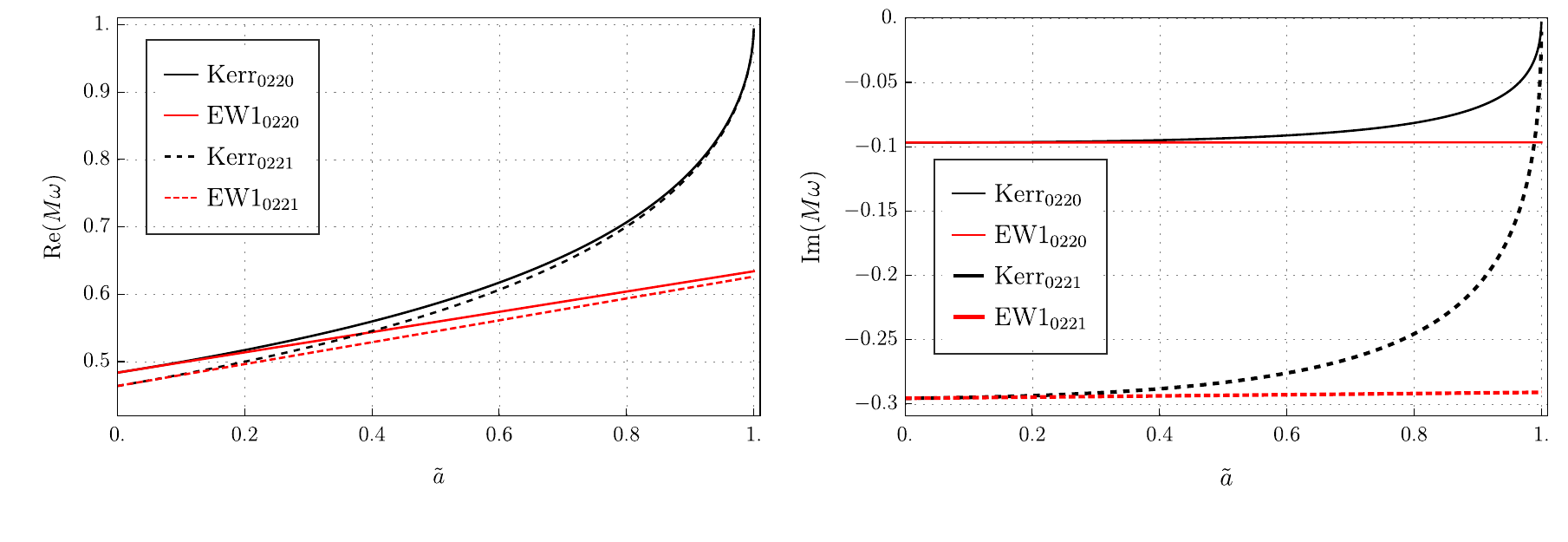}  
  \label{subfig:LinearB0_Kerr22n_Re}
\end{subfigure}
\vspace*{-0.5cm}
\caption{The real (left) and imaginary (right) parts of the $(\ell, m, n)=\{(2,2,0), (2,2,1)\}$ %(3,2,0), (4,2,0)\}$ 
scalar quasinormal mode frequencies for the Kerr solution as a function of $\tilde{a}$, compared to the corresponding modes of the linearized Ernst-Wild geometry (for $ B \rightarrow 0$). The differences for the $\{(3,2,0),(4,2,0)\}$ modes are very similar.}
\label{fig:LinearB0_Kerr}
\end{figure*}

As is apparent from Eq.~\ref{e:decoupled}, the analysis of perturbations in the Ernst-Wild spacetime simplifies considerably in the limit of small $\tilde{B}^2 \ll 1$ if we also restrict attention to slow rotation $\tilde a \ll 1$. In particular, the equations decouple if we drop corrections of ${\cal O}(\tilde{a}^2)$. Thus, we will use a second simplifying approximation.
\begin{enumerate}
   \item[(A2)] {\it Perturbative analysis at ${\cal}(\tilde{a}\tilde{B}^2)$}: A perturbative expansion in $\tilde{B}$ is well-motivated physically on the grounds that astrophysical magnetospheres are not expected to approach $BM{\sim}1$. Moreover, the large $B$ asymptotics are those of the Melvin universe which exhibits normal (as opposed to quasinormal) modes, significantly impacting the spectrum analyzed in \cite{Brito:2014super}. As described in Appendix A, the restriction to ${\cal O}(\tilde{B}^2)$ restores the Kerr-like asymptotic boundary conditions, and ensures that the QNM only reflect the impact of the magnetic field near the black hole. A perturbative expansion in $\tilde{a}$ is less well-motivated physically, as the LVK catalogue centers on remnant black holes with $\tilde{a} \sim 0.7{-}0.8$. However, when combined with a perturbative expansion in $\tilde{B}$, this approximation allows for separable perturbations, and numerically even for $\tilde{a}\sim 0.7{-}0.8$, the QNM spectrum for the linearized geometry agrees   quantitatively to $\sim$25\% with the exact Kerr modes. This comparison is illustrated in Fig.~\ref{fig:LinearB0_Kerr} (see also~\cite{Cardoso:2019mqo,Wagle:2021tam}). The fractional errors are larger for the imaginary part than the real part of the modes. To formalize the expansion, we can introduce a small dimensionless parameter $\ep$, scaling $\tilde{a}\rightarrow \tilde{a}\ep^2$ and $\tilde{B}^2\rightarrow \tilde{B}^2\ep$, and then truncate the expansion at ${\cal O}(\ep^{3})$ which retains terms of ${\cal O}(\tilde{B}^2, \tilde{a}, \tilde{a}\tilde{B}^2)$. Terms of ${\cal O}(\tilde{B}^4)$ and  ${\cal O}(\tilde{B}^6)$, while parametrically allowed under this scaling regime, are small inside the Melvin radius for the $\tilde{B}$-range considered here and are excluded to ensure Kerr-like large radius asymptotics as described above. 
\end{enumerate}

%%%%%%%%%
In combination, approximations (A1) and (A2) dramatically simplify the analysis. Under the scaling regime outlined above, at ${\cal O}(\tilde{a}\tilde{B}^2)$ the Ernst-Wild line element simplifies to the following,
\begin{equation}
    \begin{split}
    ds^2 &= \frac{1}{2r} (2 M-r)(2 + B^2 r^2 \sin ^2 \theta )dt^2 \\
        &+ \frac{r}{4M-2r}(-2 - B^2 r^2+ B^2 r^2 \cos^2 \theta )dr^2 \\
        &+ (r^2 + \frac{1}{2} B^2 r^4 \sin^2 \theta)d\theta^2 \\
        & + \frac{1}{4}r^2 (4 - B^2 r^2 + B^2 r^2 \cos 2 \theta) \sin^2 \theta\, d\phi^2 \\
        &- \frac{2 aM}{r} (2+B^2 r^2 (3 + \cos ^2 \theta)) \sin^2 \theta\, dt\, d\phi.
    \end{split}
\end{equation}
The horizon radius $r_+$ and angular momentum $J$ reduce to the Kerr expressions at ${\cal O}(\tilde{a})$, while the horizon angular velocity is corrected at ${\cal O} (\tilde{a}\tilde{B}^2$) \cite{Brito:2014super}, 
\begin{align}
 r_+ &= 2M + {\cal O}(\tilde a^2) \\
 J &= \tilde{a} M^2 + {\cal O}(\tilde a^3) \\
 M \Omega_H &= \frac{\tilde a}{4} + 2\tilde{a}\tilde B^2 + {\cal O}(\tilde{a}^3,\tilde{a}\tilde{B}^4).
\end{align}
The perturbation Eqs.~\ref{e:decoupled} now decouple and the radial equation takes the form,
\begin{align}
	\frac{d^2 \bar{R}_{\ell m}}{d r_*^2} + (\omega^2 - V_{\ell m})\bar{R}_{\ell m} = 0,
  \label{Reqn}
\end{align}
with an effective potential $V_{\ell m}$ given by 
\begin{equation}
    \label{e:Veff_EW}
    \begin{split}
    	V_{\ell m} &= \frac{4 \tilde a m r \omega  \left(2 \tilde B^2 r^2+M^2\right)}{r^4} \\
        &-  \frac{(2M-r)\left(\tilde B^2 m^2 r^3+\ell (\ell +1) M^2 r+2 M^3\right)}{ M^2\,r^4}.
    \end{split}
\end{equation}

%%%%%%%%%%%%%%%%%%%%%%%%%%%%%%%%%%%%%%%%%%%%%%%%%%%%%%%%%%%%%%%%%%%%%%%%%%%%%%%%%%%%%%%%%%
\section{QNM spectrum via continued fractions \label{sec:QNM_computation}}
%%%%%%%%%%%%%%%%%%%%%%%%%%%%%%%%%%%%%%%%%%%%%%%%%%%%%%%%%%%%%%%%%%%%%%%%%%%%%%%%%%%%%%%%%%
In this section the massless scalar quasinormal mode spectrum of the Ernst-Wild black hole will be investigated using Leaver's well-known continued fraction technique \cite{Leaver:1985analyticQNMKerr,Leaver:PhysRevD.45.4713}. 
For this purpose, it is convenient to re-express (\ref{Reqn}) 
as an equation for $R_{\ell m}(r) = \bar{R}_{\ell m}(r)/r$ in terms of the original radial coordinate, 
\begin{align}
    \label{e:radialODE}
    &\Delta \frac{d}{dr }\left ( \Delta \frac{dR_{\ell m}}{dr}\right ) 
    {+} \Biggr [ \omega ^2 r^4 {-} 4 \tilde{a} m \omega r \left(M^2{+}2 \tilde B^2 r^2\right) \\
    \nonumber
    & \qquad - \left(\ell(\ell+1) - \frac{2 M}{r} -  \frac{\tilde B^2 m^2 r^2}{M^2} \right ) \Delta \Biggr ] R_{\ell m}=0,
\end{align}
where $\De = r(r-2M)$ to $\mathcal{O}(\tilde{a} \tilde B^2)$ exhibits two regular singular points at $r =0 \text{ and at } r= 2 M$ and an irregular singular point at spatial infinity. 
The asymptotic solutions are given by
\begin{align}
    R_{\ell m}(r) \sim 
    \begin{dcases}
        (r-2 M)^{- \sqrt{2 M} \sqrt{\tilde a m \omega -2 M \Omega ^2-\rho }}, & \mathrm{\ as\ } r \rightarrow 2 M, \\
        \label{e:Rellm(r)}
       r^\frac{i \left(\rho + 4 M \Omega ^2\right)}{2 \Omega } e^{i \Omega r }, & \mathrm{\ as\ } r \rightarrow \infty,
    \end{dcases} 
\end{align}
where for convenience we have defined
\begin{subequations}
    \begin{eqnarray}
        M \Omega &\equiv&\sqrt{M^2\,\omega ^2-m^2 \tilde B^2},\\
        M \rho &\equiv& 2 \tilde B^2 (m^2-4 M \tilde a m \omega ) .
    \end{eqnarray}
\end{subequations}
We introduce a power series ansatz which incorporates the required boundary conditions,
\begin{align}
	  R_{\ell m}(r) &= r^\frac{i \left(\rho +4 M \Omega ^2\right)}{2 \Omega } r^{\sqrt{2M} \sqrt{\tilde a m \omega - 2M \Omega ^2 -\rho}} \nnl
    & \times (r-2 M)^{- \sqrt{2 M} \sqrt{\tilde a m \omega -2 M \Omega ^2-\rho }} \nnl
    \label{e:LeaverSeriesSolution}
    &\times  e^{i \Omega r } \sum_{k=0}^\infty d_n \left (\frac{r-2M}{r} \right )^k,
\end{align}
where the expansion coefficients $\{d_k\ : \ k= 1, 2, \dots, n\}$ are determined by a three-term recurrence relation starting with $d_0 = 1$,
\begin{align}
    \label{e:recurrenceRelation1}
	\alpha_0 d_1 + \beta_0 d_0 &= 0, \\
    \label{e:recurrenceRelation2}
	\alpha_n d_{n+1} + \beta_n d_{n} + \gamma_{n} d_{n-1} &= 0, \quad \mathrm{for\ } n \geq 1.
\end{align}
The recursion coefficients $\alpha_n,\, \beta_n, \text{ and } \gamma_n$ can be expressed as simple functions of $n$ and the parameters of the differential equation:
\begin{subequations}
    \begin{eqnarray}
        \alpha_n &=& n^2 + (c_0 + 1)n + c_0, \\
        \beta_n &=& - 2 n^2 + (c_1+2)n + c_3,\\
        \gamma_n &=& n^2 + (c_2-3)n + c_4 - c_2 + 2.
    \end{eqnarray}
\end{subequations}
where the intermediate constants $c_n$ are defined by 
\begin{subequations}
    \label{e:cn_Coefficients_CF}
    \begin{eqnarray}
        c_0 &=& 1-2  \sqrt{2 } \sqrt{-M \left(-{\tilde a} m \omega +2 M \Omega ^2+\rho \right)}, \\
        c_1 &=& 8 i M \Omega +\frac{i \rho }{\Omega }-2  +2 c_0,\\
        c_2 &=& -4 i M \Omega -\frac{i \rho }{\Omega }+2 - c_0,\\
        c_3 &=& 2 M \left(-\tilde a m \omega +8 M \Omega ^2+3 \rho \right)- \ell (\ell + 1) \nonumber \\
        &+& c_0 \left ( 4 i M \Omega +\frac{i \rho }{2 \Omega }-1 \right  )  ,\\
        c_4 &=&   4 i \sqrt{2} M^{3/2} \Omega  \sqrt{\tilde a m \omega -2 M \Omega ^2-\rho }\nonumber \\
        &-& \frac{i \rho  \left(-1+4 i M \Omega +\sqrt{2} \sqrt{M} \sqrt{\tilde a m \omega -2 M \Omega ^2-\rho }\right)}{\Omega } \nnl
        &+& 2 \tilde a m M \omega -\frac{\rho ^2}{4 \Omega ^2} + c_0.
    \end{eqnarray}
\end{subequations}
The recurrence relation given by Eq.~\ref{e:recurrenceRelation2} corresponds to a second-order difference equation, and there is one solution, referred to as the minimal solution \cite{Gautschi1967computational}, which leads to a convergent power series. For discrete values of $\omega = \omega_{s \ell m  n}$\footnote{The spin subscript $s$ will be dropped when considering quasinormal modes arising from scalar perturbations. } -- the quasinormal frequencies --  the series in Eq.~\ref{e:LeaverSeriesSolution} is absolutely convergent as $r \rightarrow \infty$ \cite{Leaver:1985analyticQNMKerr}. 
The ratio of successive $d_n$ is given by the following infinite continued fraction,
\begin{align}
    \frac{d_{n+1}}{d_n} &= \cfrac{- \gamma_{n+1}}{\beta_{n+1} - \cfrac{\alpha_{n+1} \gamma_{n+2}}{\beta_{n+2} - \cfrac{\alpha_{n+2} \gamma_{n+3}}{\beta_{n+3} - \cdots}}}\ ,
\end{align}
which is conveniently expressed in the following compact notation,
\begin{equation}
    \frac{d_{n+1}}{d_n} = \frac{- \gamma_{n+1}}{\beta_{n+1}\, -} \frac{\alpha_{n+1} \gamma_{n+2}}{\beta_{n+2}\, - }\frac{\alpha_{n+2} \gamma_{n+3}}{\beta_{n+3} \, -}\cdots\ . \label{cf}
\end{equation}
Setting $n=0$ in Eq.~\ref{cf} yields the following characteristic equation for the quasinormal mode frequencies, 
\begin{align}
    \label{e:continuedFractionEquation}
    \beta_0 - \frac{\alpha_0 \gamma_1}{\beta_1 -}\frac{\alpha_1 \gamma_2}{\beta_2 -}\dots \frac{\alpha_n \gamma_{n+1}}{\beta_{n+1} -}\dots = 0,
\end{align}
which can be inverted $n$ times to yield a convenient equality between a continued fraction of finite length and the remainder of infinite length:
\begin{align}
     \left [ \beta_n - \frac{\alpha_{n-1} \gamma_{n}}{\beta_{n-1} -}\dots - \frac{\alpha_{0} \gamma_{1}}{\beta_{0}} \right ]
     &=
     \left [ \frac{\alpha_{n} \gamma_{n+1}}{\beta_{n+1} -}\dots \frac{\alpha_{n+2} \gamma_{n+3}}{\beta_{n+3} - \cdots}\right ], \nonumber \\
    \label{e:CFinversion}
     &\quad \mathrm{for\ } n \geq 1.
\end{align}
Setting $n=0$ in Eq.~\ref{e:CFinversion} gives us back Eq.~\ref{e:continuedFractionEquation} provided that for $n<0,\  \alpha_n = \beta_n = \gamma_n =0$. Eq.~\ref{e:CFinversion} possesses an infinite number of roots and numerically the $n^{\mathrm{th}}$ inversion gives the $n^{\mathrm{th}}$ stable root \cite{Leaver:1985analyticQNMKerr}, i.e.\ to determine the fundamental mode, one sets $n=0$ while to compute the $n^{\mathrm{th}}$ overtone, one sets $n$ equal to the desired overtone number. 

In practice, convergence of the continued fraction slows as the imaginary part of the quasinormal mode frequency $\omega^I_\lmn$ increases. 
\begin{figure*}[t]
    \centering
    \includegraphics[width=0.8\linewidth]{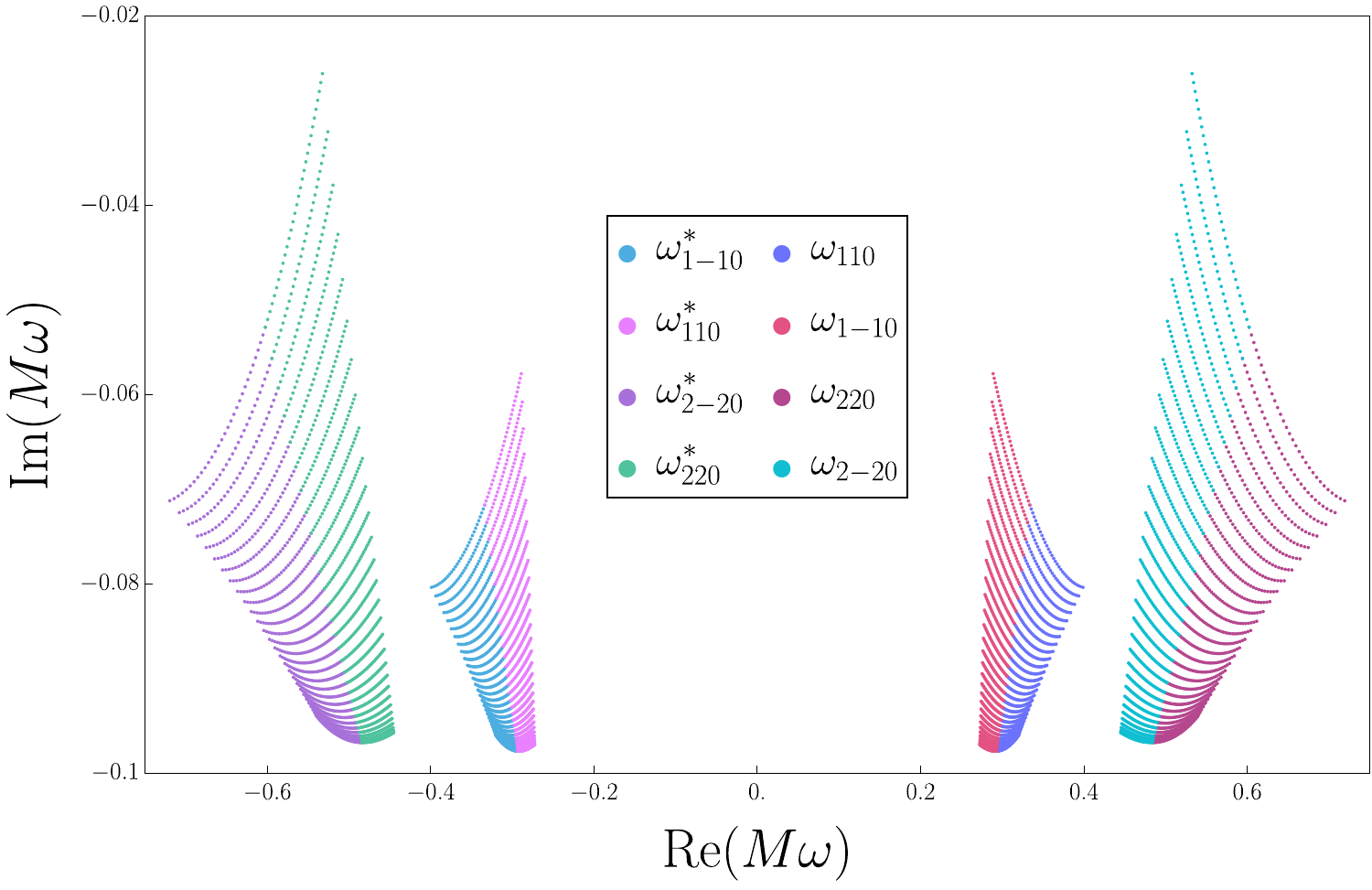}
    \caption{Quasinormal mode frequencies of the linearized Ernst-Wild geometry are presented for $0 \leq \tilde{a} , \tilde{B} \leq 0.3$ in increments of 0.01, exhibiting the $\omega_{\ell -m n} = - \omega^*_{\ell m n}$ pairing for $\ell, m = \pm 1, \pm 2$ that results from axial symmetry.  Starting from the bottom of the plots, as $\tilde a$ increases the lines move in opposite directions horizontally while as $\tilde{B}$ increases the values become less damped and move vertically upwards. These modes, including both positive and negative branches, were computed with the continued fraction methods described in Section~\ref{sec:QNM_computation}. }
    \label{fig:complexconjpairingl21m21}
\end{figure*}
Thus, we implement Nollert's technique for improved convergence \cite{Nollert:1993qnm}, first solving asymptotically for the `remainder' $R_N(\omega)$, namely the infinite continued fraction inferred from (\ref{e:CFinversion}) for $n>N\gg 1$. $R_N$ satisfies the following recurrence relation for $N\gg 1$,
\begin{align}
    R_N = - \frac{\gamma_{N+1}}{\beta_{N+1} - \alpha_{N+1} R_{N+1}}.
\end{align}
The asymptotics are readily determined from the explicit form of the recurrence relation~\ref{e:recurrenceRelation2}, and we find that, up to order $1/N$, the minimal solution satisfies 
\begin{align}
    \label{e:NollertImprovement}
     \frac{d_{N+1}}{d_N} = 1 &- \frac{2 (-1)^{3/4} \sqrt{M \Omega}}{\sqrt{N}} \nnl
     &- \frac{1}{N}\left ( \frac{3}{4} +4 i M \Omega +\frac{i \rho }{2 \Omega } \right) + {\cal O}(N^{-3/2}).
\end{align}

%%%%%%%%%%%%%%%%%%%%%%%%%%%%%%%%%%%%%%%%%%%%%%%%%%%%%%%%%%%%%%%%%%%%%%%%%%%%%%%%%%%%
\subsection{Numerical QNM results}\label{subsec:NumericalQNMs}
%%%%%%%%%%%%%%%%%%%%%%%%%%%%%%%%%%%%%%%%%%%%%%%%%%%%%%%%%%%%%%%%%%%%%%%%%%%%%%%%%%%%
The remaining task involves determining the roots of an $N$-term continued fraction given by~\ref{e:continuedFractionEquation}, whose solutions are the quasinormal modes of the desired system. For a given set of values labeled by $\lambda=\{\tilde a, \tilde B, M, \ell, m, n  \}$, the continued fraction given by Eq.~\ref{e:continuedFractionEquation} will depend only on the scalar quasinormal frequency $\omega_{\lmn} = \omega_{0 \lmn}$ along with a chosen value of $N$.  

For numerical purposes, it is convenient to normalize the frequencies on the scale $1/M$, as we observe that the mass $M$ can be removed from the perturbation equation, Eq.~(\ref{e:radialODE}), via the combined scaling $r\rightarrow Mr$ and $\om \rightarrow \om/M$. Thus we can set $M=1$ in what follows, with the understanding that the computed frequencies determine the combination $M\om$.

The equation can be solved using a root finding algorithm 
beginning with some initial guess $\omega^0_{\lmn}$.
To ensure convergence of the mode $\omega_\lmn$ we will use the technique outlined in \cite{Jaramillo:2020tuu,Siqueira:2022tbc,bohra2023gravitational} to fine-tune the required number of terms $N$ needed in the continued fraction. 
In each step we begin with the starting value of $N$ and determine $\omega_\lmn$, increasing $N$ by $\delta N$, until the relative difference between the last two adjacent values meets a specified precision threshold. That is, for a chosen set of parameters $\lambda$, if $\omega_\lmn(N)$ denotes the QNM frequency determined with an $N$-term continued fraction, the iteration will stop when 
\begin{align}
    \label{e:epislonConvergence}
    \log_{10} \left | \frac{\omega_\lmn(N + \delta N) - \omega_\lmn(N)}{\omega_\lmn(N)} \right | \Biggr |_{\lambda} < \epsilon,
\end{align}
with $\omega_\lmn = \omega_\lmn(N)$ recorded as the value for the quasinormal mode.

Earlier calculations in the literature for Kerr \cite{Leaver:1985analyticQNMKerr,Berti:2009kk} and Ernst \cite{Konoplya:2007yy,Becar:2022wcj}  black holes were used to validate our implementation of the continued fraction method. The non-Hermitian and thus inherently dissipative nature of the QNM (quasi)spectrum calculation requires the use of high machine precision throughout intermediate steps to avoid rounding errors, 
and in this work we have set the convergence criterion to $\epsilon = -7$.

We use two iterations to determine the QNM frequencies $\omega_{\lmn}$ as a function of $\tilde{a}$ and $\tilde{B}$. The first iteration begins by setting $\ell, m,  \text{ and } n$ and simultaneously  fixing $\tilde{a} =0$. The parameter space for $\tilde{B} $ is then scanned incrementally through the range $(0, 0.3)$. In this work we take the maximum value of the magnetic field to be $\tilde{B} = 0.3$, given the constraints of the linearized model. Using the Schwarzschild QNM $\omega_{\ell 0 n} = \omega^0_\lmn$ as the initial starting guess in the root finding algorithm, the parameter space of $\tilde B$ is determined in steps of $\delta \tilde{B} =0.01$ up to the point $(\tilde{a}, \tilde{B})=(0, 0.3)$. A list of values $(0, \tilde{B}, \omega_{\ell m n}, \No)$ is constructed where $\No$ is the optimal value of $N$ ensuring the convergence criterion defined in Eq.~\ref{e:epislonConvergence}. The second iteration involves using this constructed list as the seed solution while scanning $\tilde a $. More precisely, for a given value of $\tilde{B} $, the list constructed in the first iteration is fed into the root finding algorithm while incrementally varying $\tilde{a} $, starting from $\tilde{a} = 0.001$ - which corresponds to a slowly rotating Kerr black hole up to $\tilde{a} = 0.3$ in steps of $\delta \tilde{a} = 0.01$. For each value of $\tilde{B}$, the initial starting guess is given by the result for $\omega^0_{\lmn} = \omega_{\ell m n}(\tilde{a} = 0, \tilde{B})$ and the $\No$ value are used as combined starting points to speed up calculations in the second iteration while only the spin parameter is varied.  
This procedure results in a list of values  $(\tilde{a}, \tilde{B},  \omega_{\ell m n})$ for the linearized Ernst-Wild black hole, exhibited graphically for several modes in Fig.~\ref{fig:complexconjpairingl21m21}.

A notable feature is that increasing $\tilde{B}$ leads to a reduction in the imaginary part of the QNM frequency in all cases. This reflects the tendency of the pure Melvin geometry to define box-like boundary conditions and exhibit normal as opposed to quasinormal modes. Another notable feature is the symmetry between positive and negative branches of modes. An analogous symmetry was observed by Leaver in \cite{Leaver:1985analyticQNMKerr} in the case of the Kerr geometry, and can be understood in a similar way. In particular, if $\omega_{\ell m n}$ is a QNM frequency with azimuthal index $m$, then $\omega_{\ell -m n} = - \omega_{\ell m n}^*$. Similarly, the linearized EW equations contain combinations of $m^2 \tilde{B}^2$, and $m \tilde{a} \tilde{B}^2$ meaning that if $m \rightarrow -m$ then $m^2 \tilde{B}^2 $ should remain invariant while $m \tilde{a} \tilde{B}^2 \rightarrow -m \tilde{a} \tilde{B}^2$. These symmetries are apparent on computing the positive and negative branch of the quasinormal mode frequencies, as shown in Fig.~\ref{fig:complexconjpairingl21m21}.
%%%%%%%%%%%%%%%%%%%%%%%%%%%%%%%%%%%%%%%%%%%%%%%%%%%%%%%%%%%%%
\subsection{Numerical QNM interpolation}
%%%%%%%%%%%%%%%%%%%%%%%%%%%%%%%%%%%%%%%%%%%%%%%%%%%%%%%%%%%%%
\begin{figure*}[tbp]
\begin{subfigure}{\textwidth}
  \centering
  \includegraphics[width=\linewidth]{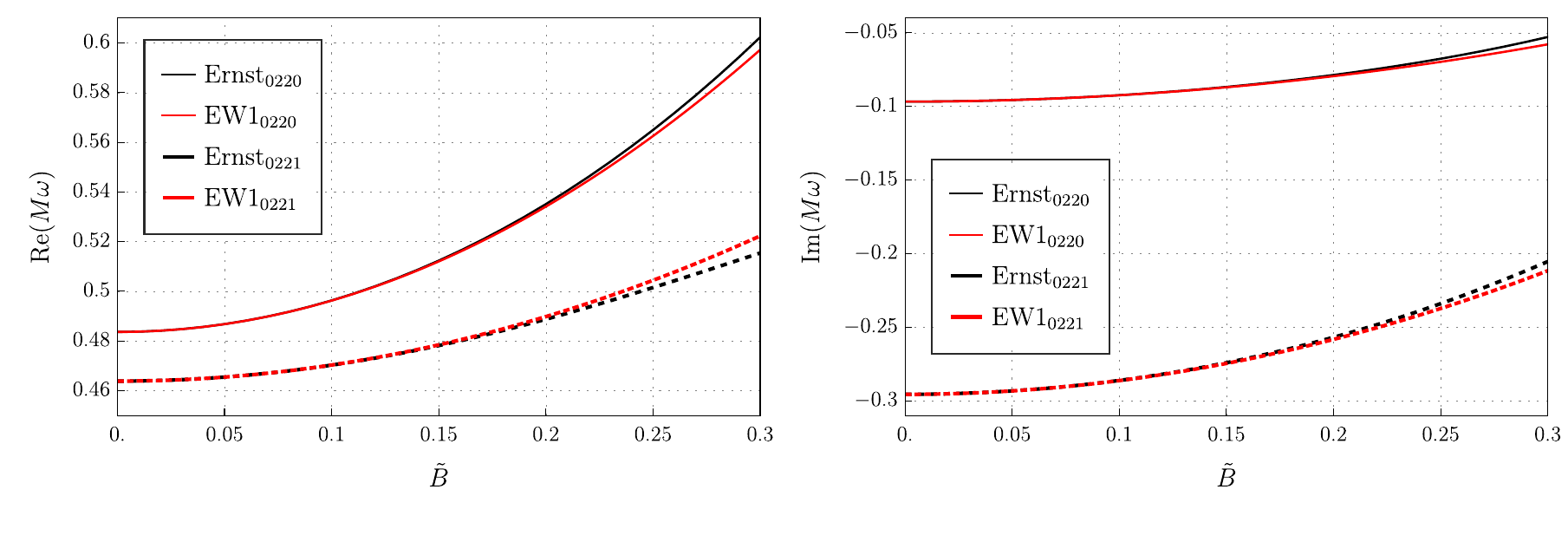}
\end{subfigure}
\vspace*{-0.5cm}
\caption{The real (left) and imaginary (right) parts of the $(\ell, m, n)=\{(2,2,0), (2,2,1)\}$ %(3,2,0), (4,2,0)\}$ 
scalar quasinormal mode frequencies for the Ernst solution as a function of $\tilde{B}$, compared to the corresponding modes within the linearized fit to Ernst-Wild spectrum in (\ref{eq:fit}) (for $\tilde a \rightarrow 0$). The differences for the $\{(3,2,0),(4,2,0)\}$ modes are very similar.}
\label{fig:ErnstvsEW1}
\end{figure*}

As outlined above, the quasinormal modes were computed over the parameter range $\tilde{a}, \tilde{B} \leq 0.3$. However gravitational wave detectors routinely see remnant spin values up to $\tilde a \approx 0.8$. Therefore, in this section we will devise a method to linearly extrapolate the results obtained for the QNMs up to $\tilde a = 1$ in order to utilize the LVK analysis tool \texttt{pyRing}.
In the limit of slow rotation and small (astrophysically relevant) magnetic fields the quasinormal frequencies ($\omega_\lmn = \omega^{\mathrm{R}}_\lmn + i \omega^{\mathrm{I}}_\lmn$) for a given $\ell, m, n$ can be expanded in the form,
\begin{equation}
    \begin{split}
        \omega_{\lmn}= \omega_{\ell 0 n} + a_\lmn \tilde{a} + b_\lmn \tilde  B^2 &+ c_\lmn \tilde{a} \tilde B ^2 \\
        &+{\mathcal O}( \tilde a^2, \tilde B^4),  
    \end{split} \label{eq:fit}
\end{equation}
where $\omega_{\ell 0 n}$ is the Schwarzchild QNM for a given $\ell \text{ and } n$ as listed in \Cref{tab:SchwarzschildQNMs}.
We compute the coefficients $a_\lmn$, $b_\lmn$, and $c_\lmn$ by taking numerical derivatives of the data obtained in the preceding section. In particular, $a_\lmn$ describe the first derivative of the Kerr spectrum in the linearized regime of $\tilde a$, $b_\lmn$ describe the first derivative of the Ernst spectrum in the linearized regime of $\tilde{B}^2$, and  $c_\lmn$ result from the numerical second derivative with respect to both $\tilde a$ and $\tilde{B}^2$. 
The results for these coefficients are tabulated in \Cref{tab:realcoefficientsposbranch,tab:imaginarycoefficientsposbranch}, with those restricted to linearized Kerr agreeing with \cite{Cardoso:2019mqo}. A comparison between the modes obtained directly from the Ernst geometry at ${\cal O}(\tilde{B}^2)$ and the coefficients $b_{lmn}$ is shown in Fig.~\ref{fig:ErnstvsEW1}.

\begin{table}[tbp]
    \caption{\label{tab:SchwarzschildQNMs}Schwarzschild QNM frequencies (with real $\omega^R_{\ell 0 n}$ and imaginary $\omega^I_{\ell 0 n}$ parts) for the fundamental ($n=0$) and first overtone ($n=1$). These results also correspond to the $m=0$ modes for Ernst and EW1 and are ordered by the multipole number $\ell$. }
    \begin{tabular}{cccccc}
    \toprule
    $\ell $ & $n$ & \hphantom{X} & $\omega^R_{\ell 0  n }$ & \hphantom{X} & $\omega^I_{\ell 0 n }$\\
    \midrule
    0 & 0 & \hphantom{X}  & 0.110455 & \hphantom{X} & -0.104896 \\
    \midrule
    \multirow{2}{*}{1} & 0 & \hphantom{X} & 0.292936 & \hphantom{X}  & -0.097660\\
        & 1  & \hphantom{X} & 0.264449 & \hphantom{X} & -0.306257  \\
    \midrule
    \multirow{2}{*}{2} & 0 & \hphantom{X} & 0.483644 & \hphantom{X} & -0.096759  \\
        & 1 & \hphantom{X} & 0.463851& \hphantom{X}  & -0.295604 \\
    \midrule
    \multirow{2}{*}{3} & 0 & \hphantom{X} & 0.675366 & \hphantom{X}  & -0.096500  \\
        &  1  & \hphantom{X} & 0.660671 & \hphantom{X}  & -0.292285 \\
    \midrule
    \multirow{2}{*}{4} & 0 & \hphantom{X} & 0.867416 & \hphantom{X}  & -0.096392  \\
        & 1 & \hphantom{X} & 0.855808 & \hphantom{X}  & -0.290876  \\[0.075cm]
    \hline
    \bottomrule
    \end{tabular}
\end{table}
\begin{table}[tbp]
    \caption{Coefficients of the linearized extrapolation for the real part of the (positive branch) EW quasinormal mode frequencies $\omega^R_\lmn$ for $\ell=1,2,3,4$ (fundamental $n=0$ mode) and $\ell=2$ (overtone $n=1$).}
    \label{tab:realcoefficientsposbranch}
    \begin{tabular}{rrrccccr}
        \toprule
        %n$ & $\ell$ & $m$ & \hphantom{X} & \multicolumn{4}{c}{} \\
        $\ell$ & $m$ & $n$ & \hphantom{X} & \multicolumn{1}{c}{$\omega^R_{\ell 0  n }$} & \multicolumn{1}{c}{$a^R_{\ell m n}$} & \multicolumn{1}{c}{$b^R_{\ell m n}$} & \multicolumn{1}{c}{$c^R_{\ell m n}$} \\
        \midrule
         \multirow{2}{*}{1} &  \multicolumn{1}{r}{$-1$} & \multirow{2}{*}{0} & \hphantom{X} & 0.292936 & -0.077158 & 0.446963 & -1.183520 \\
        %& & 0  & \hphantom{X} & $0.292936$ & $c2$ & $c3$ & $c4$ \\
        & \multicolumn{1}{r}{$1$} & & \hphantom{X} & 0.292936 & 0.077158 & 0.446963 & 1.183520 \\
        \midrule
        \multirow{4}{*}{2} & \multicolumn{1}{r}{$-2$} & \multirow{4}{*}{0} & \hphantom{X} & 0.483644 & -0.150490 & 1.262050 & -1.970010 \\
        & \multicolumn{1}{r}{$-1$} & & \hphantom{X} & 0.483644 & -0.075245 & 0.315603 & -1.256870 \\
        %& & $0$ & \hphantom{X} & $0.483644$ & $c2$ & $c3$ & $c4$ \\
        & \multicolumn{1}{r}{$1$} & & \hphantom{X} & 0.483644 & 0.075245 & 0.315603 & 1.256870 \\
        & \multicolumn{1}{r}{$2$} & & \hphantom{X} & 0.483644 & 0.150490 & 1.262050 & 1.970020 \\
        \midrule
       \multirow{4}{*}{2} & \multicolumn{1}{r}{$-2$} & \multirow{4}{*}{1} & \hphantom{X} & 0.463851 & -0.162394 & 0.649385 & -2.716630 \\
        & \multicolumn{1}{r}{$-1$} & & \hphantom{X} & 0.463851 & -0.081197 & 0.162265 & -1.406770 \\
        %& & $0$ & \hphantom{X} & $0.463851$   & $c2$ & $c3$ & $c4$ \\
        & \multicolumn{1}{r}{$1$} & & \hphantom{X} & 0.463851 & 0.081197 & 0.162265 & 1.409550 \\
        & \multicolumn{1}{r}{$2$} & & \hphantom{X} & 0.463851 & 0.162394 & 0.649385 & 2.709890 \\
        \midrule
        \multirow{6}{*}{3} & \multicolumn{1}{r}{$-3$} & \multirow{6}{*}{0} & \hphantom{X} & 0.675366 & -0.224041 & 2.122420 & -2.672430 \\
        & \multicolumn{1}{r}{$-2$} & & \hphantom{X} & 0.675366 & -0.149361 & 0.943538 & -2.281380 \\
        & \multicolumn{1}{r}{$-1$} & & \hphantom{X} & 0.675366 & -0.074680 & 0.235920 & -1.290650 \\
        %& & $0$ & \hphantom{X} & $0.675366$ & $c2$ & $c3$ & $c4$ \\
        & \multicolumn{1}{r}{$1$} & & \hphantom{X} & 0.675366 & 0.074680 & 0.235920 & 1.290660 \\
        & \multicolumn{1}{r}{$2$} & & \hphantom{X} & 0.675366 & 0.149361 & 0.943538 & 2.281410 \\
        & \multicolumn{1}{r}{$3$} & & \hphantom{X} & 0.675366 & 0.224042 & 2.122420 & 2.672480 \\
        \midrule
        \multirow{8}{*}{4} & \multicolumn{1}{r}{$-4$} & \multirow{8}{*}{0} & \hphantom{X} & 0.867416 & -0.297772 & 2.991320 & -3.345870 \\
        & \multicolumn{1}{r}{$-3$} & & \hphantom{X} & 0.867416 & -0.223329 & 1.682710 & -3.167360 \\
        & \multicolumn{1}{r}{$-2$} & & \hphantom{X} & 0.867416 & -0.148886 & 0.747899 & -2.424890 \\
        & \multicolumn{1}{r}{$-1$} & & \hphantom{X} & 0.867416 & -0.074443 & 0.186979 & -1.306440 \\
        %& & $0$ & \hphantom{X} & $0.867416$  & $c2$ & $c3$ & $c4$ \\
        & \multicolumn{1}{r}{$1$} & & \hphantom{X} & 0.867416 & 0.074443 & 0.186979 & 1.306440 \\
        & \multicolumn{1}{r}{$2$} & & \hphantom{X} & 0.867416 & 0.148886 & 0.747899 & 2.424890 \\
        & \multicolumn{1}{r}{$3$} & & \hphantom{X} & 0.867416 & 0.223330 & 1.682710 & 3.167360 \\
        & \multicolumn{1}{r}{$4$} & & \hphantom{X} & 0.867416 & 0.297773 & 2.991320 & 3.345890 \\[0.075cm]
        \hline
        \bottomrule
    \end{tabular}
\end{table}
\begin{table}[htbp]
    \caption{Coefficients of the linearized extrapolation for the imaginary part of the (positive branch) EW quasinormal mode frequencies $\omega^R_\lmn$ for $\ell=1,2,3,4$ (fundamental $n=0$ mode) and $\ell=2$ (overtone $n=1$).}
    \label{tab:imaginarycoefficientsposbranch}
    \begin{tabular}{rrrccccr}
        \toprule
        %n$ & $\ell$ & $m$ & \hphantom{X} & \multicolumn{4}{c}{} \\
        $\ell$ & $m$ & $n$ & \hphantom{X} & \multicolumn{1}{c}{$\omega^I_{\ell 0 n}$} & \multicolumn{1}{c}{$a^I_{\ell m n}$} & \multicolumn{1}{c}{$b^I_{\ell m n}$} & \multicolumn{1}{c}{$c^I_{\ell m n}$} \\
        \midrule
        \multirow{2}{*}{1} &  \multicolumn{1}{r}{$-1$} & \multirow{2}{*}{0} & \hphantom{X} & -0.097660 & -0.000336 & 0.268150 & 0.318134 \\
        %& & 0  & \hphantom{X} & $-0.09766$ & $c2$ & $c3$ & $c4$ \\
        & \multicolumn{1}{r}{$1$} & & \hphantom{X} & -0.097660 & 0.000336 & 0.268150 & -0.318132 \\
        \midrule     
        \multirow{4}{*}{2} & \multicolumn{1}{r}{$-2$} & \multirow{4}{*}{0} & \hphantom{X} & -0.096759 & -0.000143 & 0.432264 & 0.535567 \\
        & \multicolumn{1}{r}{$-1$} & & \hphantom{X} & -0.096759 & -0.000072 & 0.108133 & 0.132246 \\
        % & & $0$ & \hphantom{X} & $-0.0967588$ & $c2$ & $c3$ & $c4$ \\
        & \multicolumn{1}{r}{$1$} & & \hphantom{X} & -0.096759 & 0.000072 & 0.108133 & -0.132246 \\
        & \multicolumn{1}{r}{$2$} & & \hphantom{X} & -0.096759 & 0.000143 & 0.432264 & -0.535561 \\
        \midrule
        \multirow{4}{*}{2} & \multicolumn{1}{r}{$-2$} & \multirow{4}{*}{1} & \hphantom{X} & -0.295604 & -0.004641 & 0.933283 & 1.131990 \\
        & \multicolumn{1}{r}{$-1$} & & \hphantom{X} & -0.295604 & -0.002321 & 0.233458 & 0.322617 \\
        %& & $0$ & \hphantom{X} & $-0.295604$   & $c2$ & $c3$ & $c4$ \\
        & \multicolumn{1}{r}{$1$} & & \hphantom{X} & -0.295604 & 0.002321 & 0.233458 & -0.322418 \\
        & \multicolumn{1}{r}{$2$} & & \hphantom{X} & -0.295604 & 0.004642 & 0.933283 & -1.132600 \\
        \midrule
        \multirow{6}{*}{3} & \multicolumn{1}{r}{$-3$} & \multirow{6}{*}{0} & \hphantom{X} & -0.096500 & -0.000078 & 0.512260 & 0.653273 \\
        & \multicolumn{1}{r}{$-2$} & & \hphantom{X} & -0.096500 & -0.000052 & 0.227786 & 0.263341 \\
        & \multicolumn{1}{r}{$-1$} & & \hphantom{X} & -0.096500 & -0.000026 & 0.056964 & 0.080038 \\
        % & & $0$ & \hphantom{X} & $-0.0964996$ & $c2$ & $c3$ & $c4$ \\
        & \multicolumn{1}{r}{$1$} & & \hphantom{X} & -0.096500 & 0.000026 & 0.056964 & -0.080027 \\
        & \multicolumn{1}{r}{$2$} & & \hphantom{X} & -0.096500 & 0.000052 & 0.227786 & -0.263310 \\
        & \multicolumn{1}{r}{$3$} & & \hphantom{X} & -0.096500 & 0.000078 & 0.512260 & -0.653221 \\
        \midrule
        \multirow{8}{*}{4} & \multicolumn{1}{r}{$-4$} & \multirow{8}{*}{0} & \hphantom{X} & -0.096392 & -0.000048 & 0.558610 & 0.726651 \\
        & \multicolumn{1}{r}{$-3$} & & \hphantom{X} & -0.096392 & -0.000036 & 0.314251 & 0.370947 \\
        & \multicolumn{1}{r}{$-2$} & & \hphantom{X} & -0.096392 & -0.000024 & 0.139677 & 0.164431 \\
        & \multicolumn{1}{r}{$-1$} & & \hphantom{X} & -0.096392 & -0.000012 & 0.034921 & 0.057358 \\
        %& & $0$ & \hphantom{X} & $-0.0964996$ & $c2$ & $c3$ & $c4$ \\
        & \multicolumn{1}{r}{$1$} & & \hphantom{X} & -0.096392 & 0.000012 & 0.034921 & -0.057357 \\
        & \multicolumn{1}{r}{$2$} & & \hphantom{X} & -0.096392 & 0.000024 & 0.139677 & -0.164431 \\
        & \multicolumn{1}{r}{$3$} & & \hphantom{X} & -0.096392 & 0.000037 & 0.314251 & -0.370944 \\
        & \multicolumn{1}{r}{$4$} & & \hphantom{X} & -0.096392 & 0.000049 & 0.558610 & -0.726644 \\[0.075cm]
        \hline
        \bottomrule
    \end{tabular}
\end{table}

Presently, there is some debate within the gravitational wave literature regarding which modes can be reliably extracted from the black hole ringdown. Indeed, it is clear that the process of ringdown following a physical binary merger is complex, and nonlinear effects may be as large as higher-order QNM contributions. In \cite{Baibhav:2023clw}, it is argued that even in the ideal situation when a binary black hole system collides head-on, only the modes $\{(2,2,0), (3,2,0), (4,2,0)\}$ significantly contribute to the strain. Reliable extraction of the first overtone $(2,2,1)$ would then depend on a high-precision determination of the longer-lived fundamental modes of subdominant multipoles. In the non-ideal case when the binary does not collide head-on, it becomes significantly more difficult to extract overtones and the dominant modes excited could differ from $\{(2,2,0), (3,2,0), (4,2,0), (2,2,1)\}$. For this reason we include the fit coefficients for the fundamental modes from $1 \leq \ell \leq4$ as well as the first overtone for $\ell=2$ to cover the range expected to be relevant in a range of ringdown analyses. 

%%%%%%%%%%%%%%%%%%%%%%%%%%%%%%%%%%%%%%%%%%%%%%%%%%%%%%
\section{\label{sec:Ringdown}Ringdown template and comparison to LVK data}
%%%%%%%%%%%%%%%%%%%%%%%%%%%%%%%%%%%%%%%%%%%%%%%%%%%%%%
In this section we utilize our QNM results for the linearized EW geometry to build and test a ringdown template against the existing LVK data archive of binary black hole mergers. This serves as a first step in modifying the usual Kerr template to model the post-merger phase of a binary black hole coalescence in the presence of an external asymptotically uniform magnetic field. We model our approach on \cite{Carullo:2021oxn}, which considered the ringdown phase of black holes in the presence of a remnant $U(1)$ charge. However, we emphasize that this is a scalar analogue template and only precise for small $\tilde{B}$. Thus it cannot be used to constrain the magnetic field scale from data. Our exploratory approach will simply be to study $B$ as a nuisance-type parameter, and the template will be obtained by simply extrapolating the perturbative QNM spectrum.

To investigate the magnetized Ernst-Wild hypothesis in the current LVK data archive, we make use of \texttt{pyRing} \cite{pyRing, Carullo:2019flw, Isi_2019}, a \texttt{python} software package specifically designed for the estimation of black hole ringdown parameters. \texttt{pyRing} implements a Bayesian analysis approach, described within the LVK analysis pipeline in Refs.~\cite{LIGOScientific:2019hgc, LIGOScientific:2020tif, Isi:2021iql, Thrane_2019,Dreyer:2003bv,Gennari:2023gmx}.  

For the remainder of this section, $ \omega_{-2 \lmn}$ will be used to denote tensor quasinormal modes and $\omega_\lmn$ will denote the scalar modes considered in the previous section.

\subsection{Constructing the ringdown template} 
Following the merger of two black holes, the newly formed remnant relaxes to a stationary state, with its late time transient expected to `ring-down' according to a superposition of the leading QNMs of the stationary final state. In practice, this waveform model comprising a superposition of damped sinusoids is only expected to hold well away from the peak of the gravitational wave signal, and nonlinear effects can compete with subleading linear modes.

To construct a waveform model for a black hole in the presence of an external magnetic field, we start from the standard Kerr template \cite{Carullo:2021oxn,Berti:2005ys,Lim:2019xrb}:
\begin{align}
    h_+ - i h_\times = \frac{M_f}{D_L}\, \sum_{\ell = 2}^\infty\, \sum_{m = - \ell}^{\ell}\, \sum_{n = 0}^{\infty} \, \left [h_{\lmn}^+ + h_{\lmn}^-\right ], 
    \label{temp}
\end{align}
where
\begin{align}
    h_{\lmn}^+(t) &= \mathcal{A}_{\lmn}^+ S_{\lmn} (\iota, \varphi)e^{-i (t-t_0)\omega_{s \lmn} + i \phi_{\lmn}^+ }\\
    h_{\ell m n}^-(t) &= \mathcal{A}_{\lmn}^- S_{\lmn}^* (\pi - \iota, \varphi)e^{i (t-t_0)\omega_{s \lmn}^* + i \phi_{\lmn}^- }.
\end{align}
Here, $\tilde \omega_{s \lmn}$ are the complex tensor ($s=-2$) ringdown frequencies of the Kerr black hole determined by the dimensionless final spin $\tilde a_f$ and the detector frame final mass given by $M_f$ defined in \cite{LIGOScientific:2016vlm}. $M_f$ is related to the source frame mass by a factor of $(1+z)$, and $D_L$ is the luminosity distance.
The $f$ subscript on parameters in this section indicates that these values refer to the final black hole of the merger, commonly referred to as the remnant. 

The amplitudes $\mathcal{A}_{\lmn}^{+/-}$ and phases $\phi_{\lmn}^{+/-}$ characterise the excitation of each mode and are inferred from the data. The inclination of the black hole spin relative to the observer's line of sight is given by $\iota$, and $\varphi$ denotes the azimuthal angle of the line of sight in the frame of the black hole, and without loss of generality is usually set to zero \cite{Carullo:2021oxn}. 
$S_{\lmn}$ are the spin-weighted spherical harmonics \cite{Berti:2014fga} and $t_0$ is the reference start time. 
The oscillation frequencies and decay times are given by $ \omega^R_{-2 \lmn} = 2 \pi f_{-2 \lmn}$ and $ \omega^I_{-2 \lmn} = 1/ \tau_{-2 \lmn}$ respectively. 
The strain time series measured at the gravitational wave detector is a linear combination of both the plus ($h_+$) and cross ($h_\times$) polarizations \cite{Isi:2021iql}, defined by
\begin{align}
    h(t) = F_+ h_+(t) + F_\times h_\times(t),
\end{align}
where $F_+= F_+(\alpha, \delta, \psi)$ and $F_\times= F_\times(\alpha, \delta, \psi)$ are the detector response functions characterized by the two polarizations \cite{mishra2023bounds}. Both of these functions depend on the parameters $(\alpha, \delta, \psi)$ which are the sky localization parameters: right ascension, declination and polarization angle respectively. 
The sky location is fixed to coincide with the maximum likelihood value determined from the full Inspiral-Merger-Ringdown (IMR) analysis \cite{LIGOScientific:2021sio}.
These detector values can be found for any ringdown event in the corresponding LVK configuration file \cite{LIGO_data_2021_5172704, LIGO_data_2022_7007370}. 

By restricting this template to a superposition of the quadrupolar fundamental mode and its first corresponding overtone, that is $(\ell,m,n) = \{(2,2,0),(2,2,1)\}$, all the amplitudes and phases will be considered as independent parameters of the fit and will be left free to vary. 
We refer to the template constructed in this manner using the Kerr QNM frequencies as Kerr$_{221}$.

To incorporate the presence of the external magnetic field $\tilde{B}_f$ into the template, we need to account for our two primary approximations A1 and A2, and accordingly model the ringdown by replacing the tensor quasinormal frequencies $\omega_{-2 \lmn}$ of the Kerr geometry with the scalar linearized Ernst-Wild quasinormal modes $\omega_\lmn$, as determined in Section~\ref{sec:QNM_computation}, which are now functions of $(M_f, \tilde{a}_f, \tilde{B}_f)$,
\be
 \omega_{-2 \lmn}|_{\text{Kerr}} \longrightarrow \omega_{0 \lmn}|_{\text{EW1}}.
 \label{sub}
\ee
It is clear that this will not allow for an actual detection of a magnetosphere, but it will allow us to test for the impact of the magnetic field as a `nuisance' parameter in the analysis. In practice, given that astrophysical magnetic fields are expected to be orders of magnitude smaller than $B_{M}$, a similar methodology may be the most practical approach even if tensor QNMs for EW were available. 
The modified template, obtained from Eq.~\ref{temp} by simply implementing the substitution Eq.~\ref{sub} will be used in the remainder of this work and labeled EW1$_{221}$ to denote linearized Ernst-Wild. 

We conclude this subsection with a reminder of some additional caveats to the analysis. The restriction of Kerr$_{221}$ or EW$_{221}$ to just a superposition of the fundamental QNM and its first overtone is an assumption. 
Although contributions from the overtones close to the peak have not been shown to correspond to any physical vibrational frequencies of the underlying geometry \cite{Baibhav:2023clw}, their usage here will be to facilitate fitting the waveform close to the signal peak for sufficient accuracy due to the low signal-to-noise ratio (SNR) in many ringdown signals \cite{mishra2023bounds} observed by LVK. The inclusion of the overtone enables the analysis to be pushed to an earlier start time, capturing more of the power which is present near the peak of the signal. This is possible since the overtones are necessarily more damped than their fundamental counterparts, and can serve to effectively remove non-linearities present near the waveform peak, albeit at the risk of potentially removing interesting contributions due to the merger and thus `overfitting' the data \cite{Baibhav:2023clw}. A second point to note is that in ringdown-only analyses, the precise value of the ringdown start time $t_0$ can have a non-negligible effect on the analysis and usually requires marginalization over its uncertainty. However, since our analysis will involve a fixed set of modes, where the primary goal is simply to consider whether the remnant is magnetized or not, the impact of the start time should be much smaller and, as in \cite{mishra2023bounds}, we will fix the start time to reduce the computational cost.

\begin{figure*}[htbp]
\begin{subfigure}{.49\textwidth}
  \centering
  % include first image
  \includegraphics[width=\linewidth]{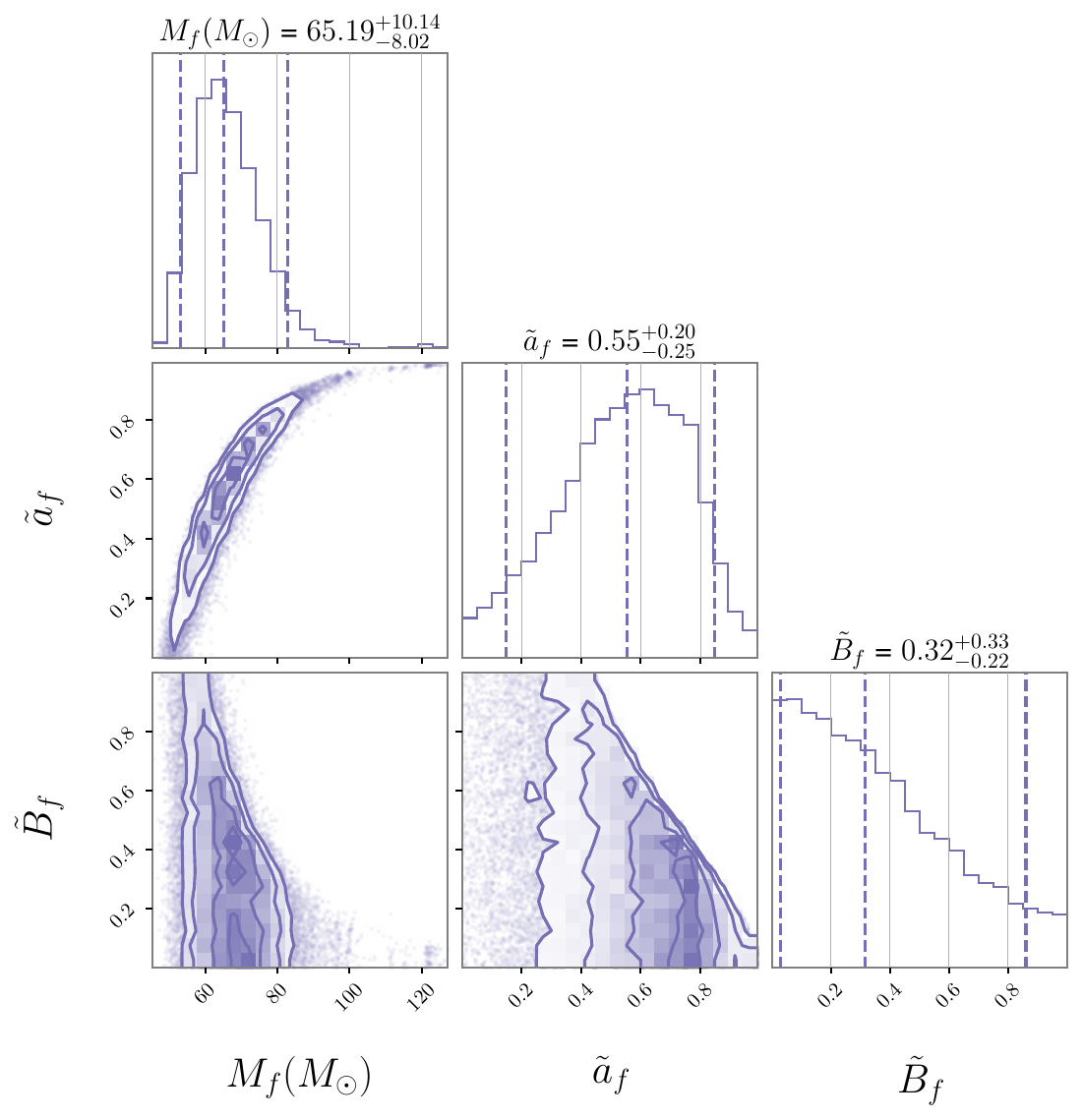}  
  \caption{GW150914}
  \label{fig:sub-first}
\end{subfigure}
\begin{subfigure}{.49\textwidth}
  \centering
  % include second image
  \includegraphics[width=\linewidth]{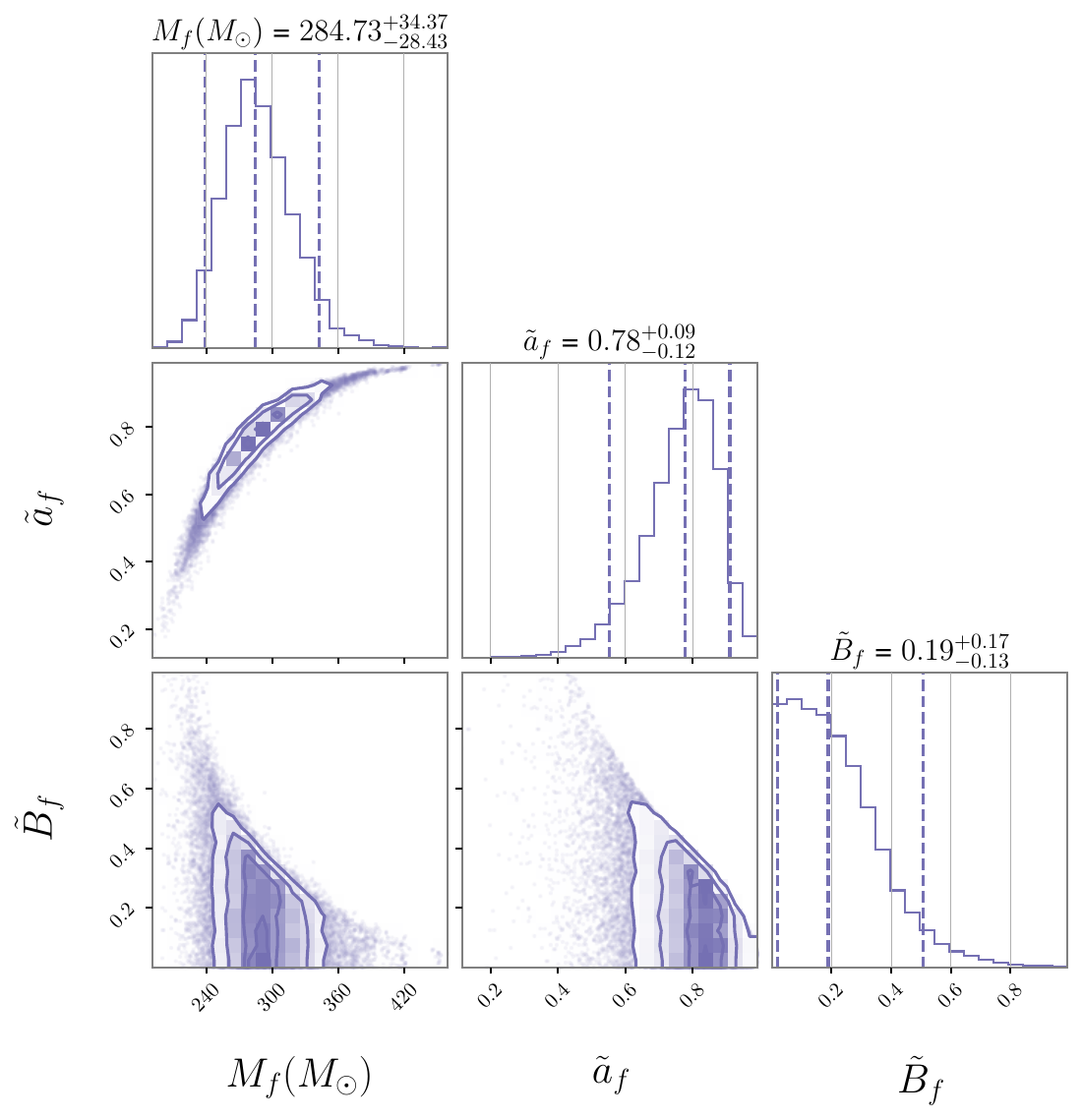}  
  \caption{GW190521} 
  \label{fig:sub-second}
\end{subfigure}
\begin{subfigure}{.49\textwidth}
  \centering
  % include third image
  \includegraphics[width=\linewidth]{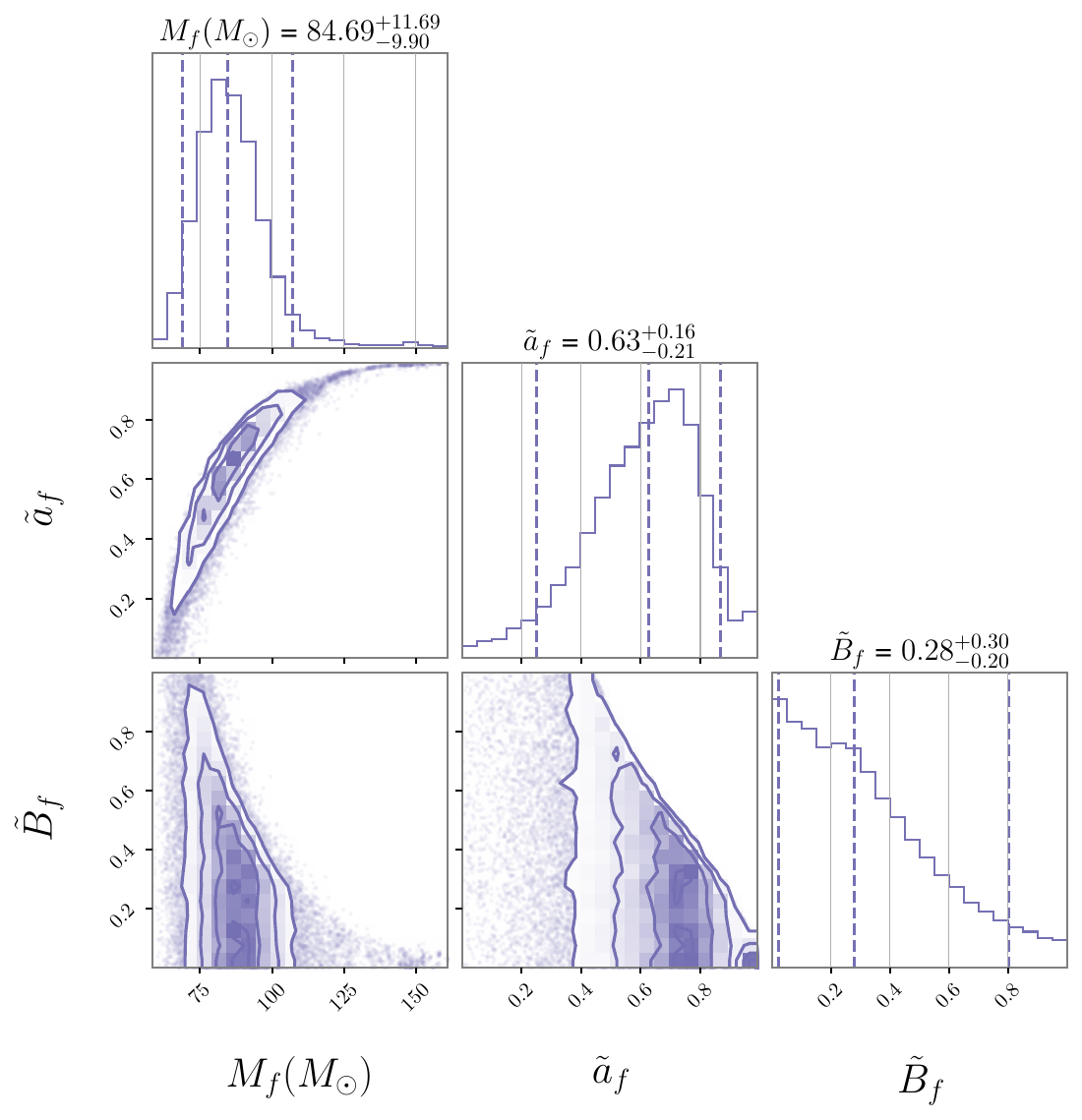}  
  \caption{GW190521\_074359}
  \label{fig:sub-third}
\end{subfigure}
\begin{subfigure}{.49\textwidth}
  \centering
  % include fourth image
  \includegraphics[width=\linewidth]{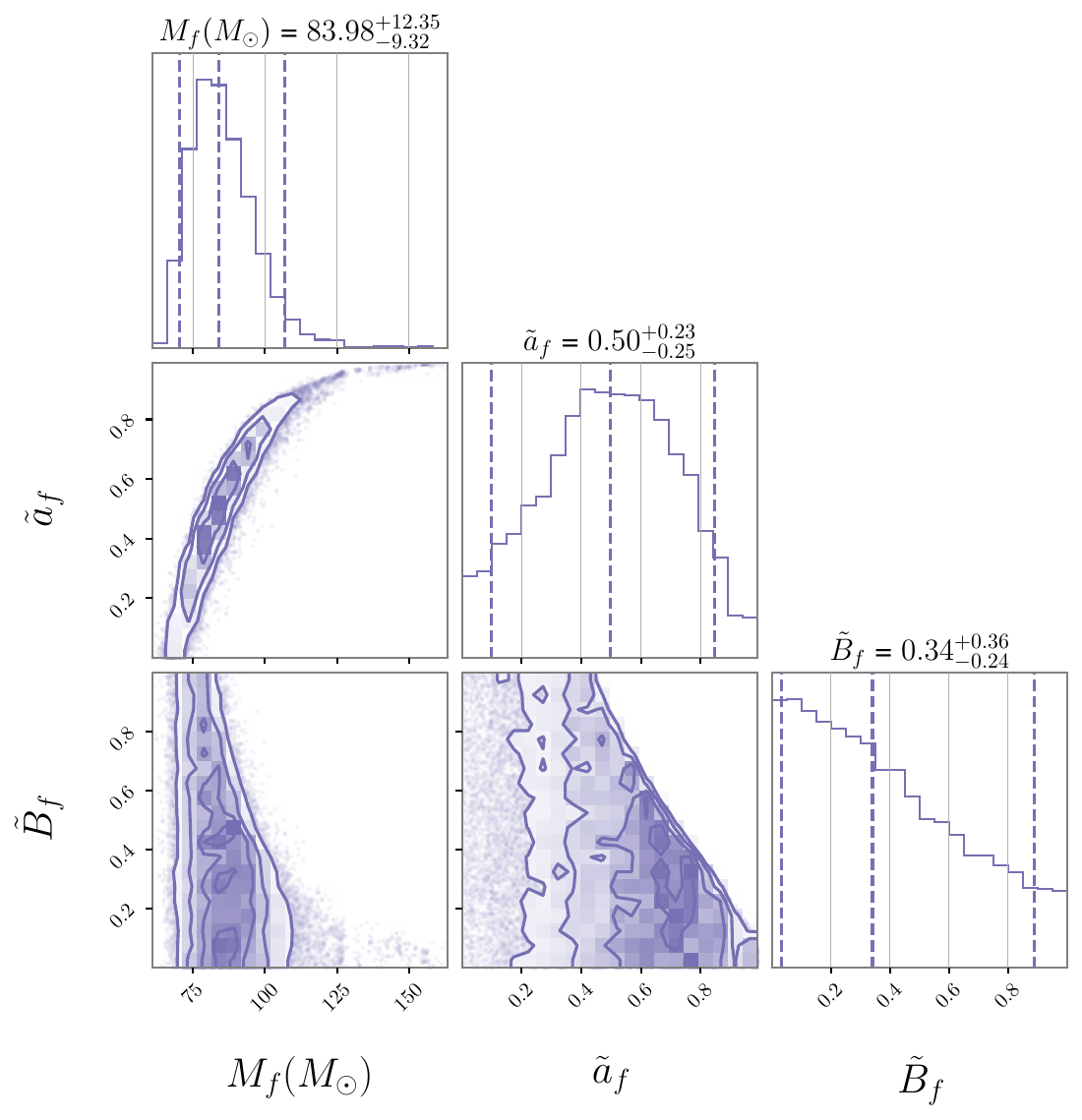}  
  \caption{GW200224\_222234}
  \label{fig:sub-fourth}
\end{subfigure}
\caption{Posterior distributions for the remnant mass $M_f$, spin $\tilde{a}_f$, and magnetic field $\tilde{B}_f$ resulting from the analysis of four gravitational wave signals of binary mergers from the LVK catalogue: GW150914, GW190521\_074359, GW190727\_060333 and GW200224\_222234. The upper and lower limits on the estimated parameters bound the $90\%$ credible interval and are shown with the dashed purple lines. 
}
\label{fig:corner_plot_pyRing}
\end{figure*}

\subsection{Time-domain likelihood}
The analysis goal is to construct the posterior distribution $p ( \theta | d )$, where $\theta$ is the set of model parameters and $d$ is the data associated with the measurement. Bayes' theorem determines $p(\theta | d ) = \mathcal L ( d | \theta)\, \pi ( \theta)/\mathcal{Z}$
in terms of $\mathcal L (d | \theta)$, the likelihood function of the data given the parameters $\theta$, while $\pi(\theta)$ is the prior distribution for $\theta$ and $\mathcal{Z} \equiv \int d \theta\, \mathcal{L}(d | \theta)\, \pi(\theta) $ is the normalization factor, referred to as the `evidence'. 

As in \cite{Carullo:2021oxn}, the likelihood will be assumed to follow a zero-mean Gaussian distribution with a standard deviation equal to the numerical uncertainty along with uniform priors on the coefficients of the template. The underlying stochastic sampling used to calculate all the posterior distributions in \texttt{pyRing} is performed using a nested sampling algorithm \cite{Skilling:2004pqw} implemented through the \texttt{python} package \texttt{CPNest} \cite{cpnest}. Each event sampler settings used in each event was 4096 live points, 4096 as a maximum number of Markov-Chain Montecarlo internal steps and a pool-size composed of 100 walkers. Increasing the live points and internal steps to 8192 did not impact the final results obtained. When the estimated precision on the logarithm of the evidence is within 0.1, the nested sampler stops.
Once these posterior samples have been generated it is then possible to generate the marginalized posteriors for any subset of the parameters by simply selecting the corresponding samples - this property of marginalized distributions can be used to visualize the output of these samples by constructing corner plots, which show the marginalized one-dimensional and two-dimensional posterior distributions for each of the parameters.

%%%%%%%%%%%%%%%%%%%%%%%%%%%%%%
\subsection{Ringdown analysis and discussion}
%%%%%%%%%%%%%%%%%%%%%%%%%%%%%%

\begin{table*}[tbp]
    \caption{Table of values comparing the IMR results for Kerr$_{221}$ \cite{KAGRA:2023pio,LIGOScientific:2019lzm,LIGOScientific:2020tif,LIGOScientific:2021sio} with the EW1$_{221}$ results for the redshifted final mass $ M_f $, dimensionless final spin $\tilde{a}_f $ and dimensionless magnetic field $\tilde{B}_f$, where the latter can be restored to physical units using Eq.~\ref{e:weakly_magntized}. For $M_f$ and $\tilde{a}_f$, we report the median value along with the range of the $90\%$ credible interval \cite{LIGOScientific:2013yzb}. For the magnetic field $\tilde{B}_f$, we report the upper limit of the $90 \%$ credible interval.
    }
    \label{tab:eventlist}
    \begin{tabular}{lllllllllc}
    \toprule
        Event & \multicolumn{3}{c}{Redshifted final mass} & \hphantom{X} & \multicolumn{3}{c}{Final spin} & \hphantom{X} & \multicolumn{1}{c}{Magnetic field} 
        \\
        & \multicolumn{3}{c}{$M_f \, [M_{\odot}]$} & \hphantom{X} & \multicolumn{3}{c}{$\tilde a_f$} 
        & \hphantom{X} & \multicolumn{1}{c}{$\tilde{B}_f$} 
        %& 
        \\
        \cmidrule{2-4}
        \cmidrule{6-8}
        \cmidrule{10-10}
        & IMR & $\mathrm{Kerr_{221}}$ & $\mathrm{EW1_{221}}$ & \hphantom{X} & IMR &  $\mathrm{Kerr_{221}}$ & $\mathrm{EW1_{221}}$ & \hphantom{X} & \multicolumn{1}{c}{$\mathrm{EW1_{221}}$} \\
        \midrule
        GW150914 &
        $ 68.8^{+ 3.6 }_{- 3.1 } $ &
        $ 71.7^{+ 13.2 }_{- 12.5 } $ &
        $ 65.2^{+ 10.1}_{- 8.0}$ &
        
        \hphantom{X} &
        $ 0.69^{+ 0.05 }_{- 0.04 } $ &
        $ 0.69^{+ 0.18 }_{- 0.36 } $ &
        $ 0.55^{+ 0.20}_{- 0.25} $ & 
        \hphantom{X} & 
        % $ 0.37^{+ 0.43}_{- 0.33} $ \\[0.075cm]
        %$ 0.32^{+ 0.33}_{- 0.22} $\\[0.075cm]
        $\leq 0.65$\\[0.075cm]

        GW190521 &
        $ 256.6 ^{+ 36.6 }_{- 30.4 } $ &
        $ 284.0 ^{+ 40.4 }_{- 43.9 } $ &
        $ 284.7 ^{+ 34.4 }_{- 28.4} $ &
        
        \hphantom{X} &
        $ 0.71 ^{+0.12 }_{- 0.16 } $ &
        $ 0.78 ^{+ 0.10  }_{- 0.22 } $ &
        $ 0.78 ^{+ 0.09}_{- 0.12 } $ &
        
        \hphantom{X} & 
         % $ 0.29^{+ 0.35}_{- 0.26} $ \\[0.075cm]
         $\leq 0.36$\\[0.075cm]

        GW190521\_074359 &
        $ 88.1^{+ 4.3 }_{- 4.9 } $ &
        $ 86.4^{+ 14.1 }_{- 14.8 } $ &
        $ 84.7 ^{+ 11.7 }_{- 9.9 } $ &
        
        \hphantom{X} &
        $ 0.72^{+ 0.05 }_{- 0.07 } $ &
        $ 0.67^{+ 0.17 }_{- 0.34 } $ &
        $ 0.63^{+ 0.16 }_{- 0.23 } $ &
        
        \hphantom{X} & 
         % $ 0.34 ^{+ 0.42}_{- 0.31} $ \\[0.075cm]
         $ \leq 0.58 $ \\[0.075cm]

        GW200224\_222234 &
        $ 90.3^{+ 6.4 }_{- 6.3 } $ &
        $ 88.6^{+ 15.5 }_{- 15.2 } $ &
        $ 84.0 ^{+ 12.4 }_{- 9.3 } $ &
        
        \hphantom{X} &
        $ 0.73^{+ 0.06 }_{- 0.07 } $ &
        $ 0.60^{+ 0.23 }_{- 0.42 } $ &
        $ 0.50 ^{+ 0.23 }_{- 0.25 } $ &
        
        \hphantom{X} & 
        % $ 0.37^{+ 0.44}_{- 0.34} $ \\[0.075cm]
        $ \leq 0.70 $ \\[0.075cm]
        \hline
        \bottomrule
    \end{tabular}
\end{table*}

\begin{figure}[tbp]
    \centering
    \includegraphics[width=\linewidth]{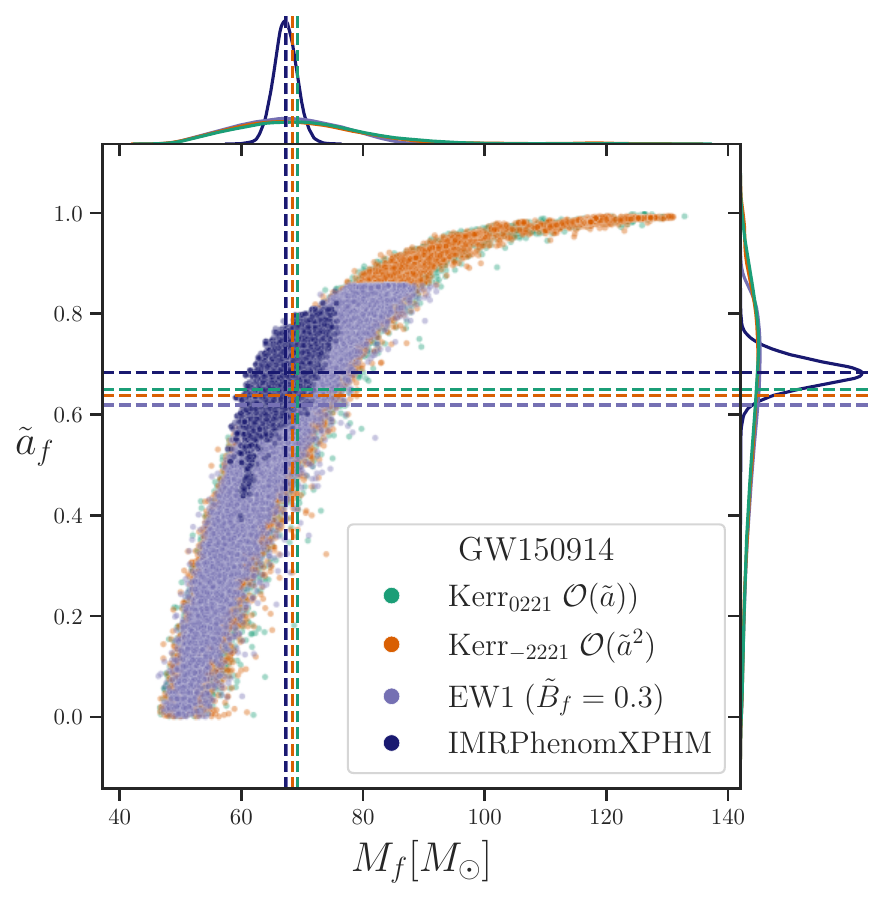}
    \caption{Posterior distribution for the remnant mass and spin resulting from the analysis of GW150914 for three different ringdown templates and the results from the full LVK IMR model using IMRPhenomXPHM (obtained from the data release \cite{ligo_scientific_collaboration_and_virgo_2022_6513631}).}
    \label{fig:GW150914_comparison}
\end{figure}
\begin{figure}
    \centering
    \includegraphics[width=\linewidth]{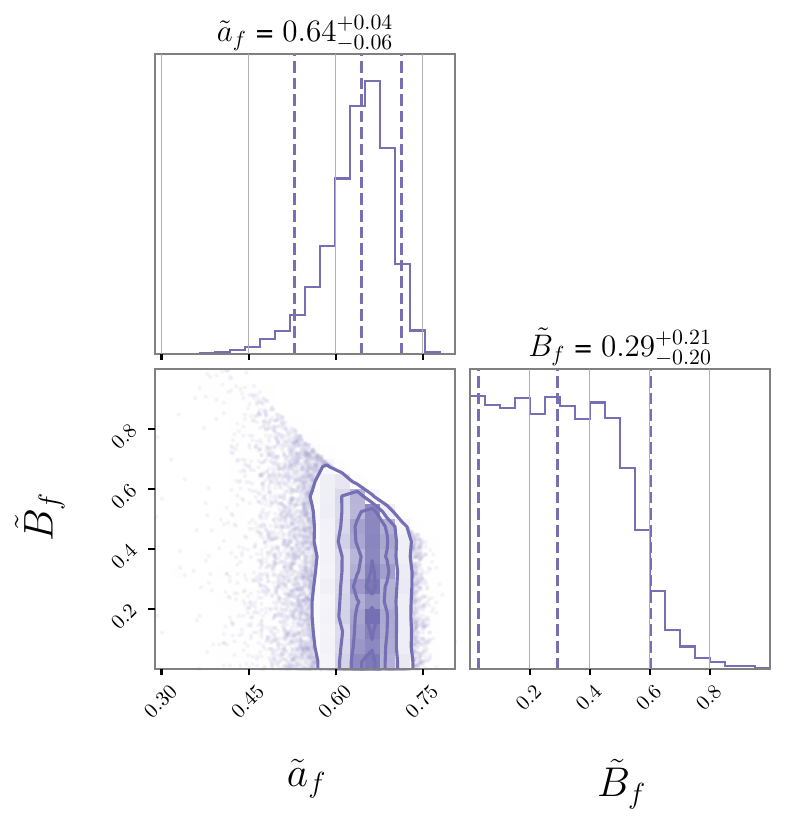}
    \caption{Posterior distributions for the remnant spin $\tilde a_f$ and magnetic field $\tilde B_f$ resulting from the anlalysis keeping the remnant mass fixed to the IMR median value: $M_f = 68.8 M_\odot$. Due to the restriction on the remnant mass, there is enhanced sensitivity to $\tilde B_f$.}
    \label{fig:GW150914_afBf_fixedMf}
\end{figure}

Having described the essential tools of our methodology, we now shift to analyzing LVK data provided by the GW Open Science Center \cite{KAGRA:2023pio,LIGOScientific:2019lzm}. Employing the described framework, and assuming that the remnant object to be a EW1 remnant described by scalar quasinormal modes, we conduct an analysis of the ringdown data for four binary black hole merger events from the LVK catalogue. The events were chosen based on their high signal to noise and sufficiently long waveform for a ringdown-only analysis. 

We initially adopt simple uniform prior distributions for the remnant mass, spin and magnetic field, $M_f \in [10,500] M_{\odot}$, $\tilde a_f \in [0,0.99]$, and $\tilde{B}_f \in [0,1]$ respectively. Note that we are choosing to extrapolate the linearized analysis to form the EW1 template over the full range of spin and magnetic field. It is clear that the extrapolation has limited applicability as one approaches large values of $a$ and $B$, but this will be sufficient to explore sensitvity to $B$ as a nuisance parameter. Finally, we also 
account for the choice $q = -2aMB$ to ensure overall electrical neutrality, we address the extremality bound $q^2 + a^2 \leq M^2$ by imposing an \textit{a priori} joint limit on the spin and magnetic field parameters $\tilde a^2_f+ 4 \tilde a_f ^2 \tilde B_f^2 <0.99$, excluding near-extremal black hole configurations. 

We also consider a more constraining choice of priors, referred to as the IMR prior \cite{Laghi:2020rgl}, with uniform distributions of $M_f$ and $\tilde a_f$ containing the $99\%$ credible intervals of their one-dimensional probability density functions determined from a full LVK IMR analysis of the event. For GW150914 this set of priors corresponds to: $M_f \in [62,75] M_\odot$ and $\tilde a_f \in [0.55,0.75]$. This choice assumes that the nuisance parameter $\tilde B$ is weakly correlated with other black hole parameters, and that the IMR gravitational wave emission follows the Kerr template. This approach naturally provides a more stringent constraint on the value of $\tilde B_f$.

The results are shown in Figs.~\ref{fig:corner_plot_pyRing}, \ref{fig:GW150914_comparison} and \ref{fig:GW150914_afBf_fixedMf}. The corner plots show in Figs.~\ref{fig:corner_plot_pyRing} exhibit the $90\%$ credible interval of the two-dimensional posterior distributions on remnant redshifted mass, spin and magnetic field for the set of four selected events using ths simple uniform priors. The results of which are also tabulated in Table~\ref{tab:eventlist}, where they are compared to results determined by LVK using the standard Kerr template. Note that care is needed in comparing the inferences from different templates. Kerr$_{221}$ and EW1$_{221}$ apply just to the ringdown signal from \texttt{pyRing}, while IMR is the full Kerr template applied also to the inspiral and merger. The inferred central values of mass and spin therefore differ, but the results for Kerr$_{221}$ and EW1$_{221}$ agree with each other and with IMR within their respective 90\% confidence intervals.

In Fig.~\ref{fig:GW150914_comparison}, we show the two-dimensional posterior for the event GW150914 to compare how different models impact the extraction of the remnant spin $\tilde{a}_f$ and mass $M_f$. The chosen models for comparison include: scalar QNM for linearized Kerr at $\mathcal O(\tilde{a})$, tensor QNM for Kerr at $\mathcal O (\tilde{a}^2)$, scalar QNM for linearized Ernst-Wild at a chosen value of $\tilde{B}_f = 0.3$ and the full LVK inspiral-merger-ringdown (IMR) posterior determined using the IMRPhenomXPHM model \cite{Pratten_2021}. Finally, Fig.~\ref{fig:GW150914_afBf_fixedMf} shows the corner plots for the more restricted priors in which the remnant mass is fixed for GW150914 to its central value from the IMR analysis. This naturally leads to a stronger constraint on $\tilde B$. We find that further restricting $\tilde a$ to its IMR-inferred value does not to a significantly stronger constraint.

It is important to emphasize that, due to the extrapolation of the linearized EW1 template, the actual constraints on $\tilde B$ are properly understood as indicative of the level of sensitivity to $\tilde B$ as a nuisance parameter, and not a true constraint on the magnetosphere. The results nonetheless indicate no evidence for a deviation from the Kerr hypothesis, and while the ensuing bounds are relatively weak, the current data already provides sensitivity to magnetic fields above $\tilde{B}\sim 0.3$. This is still very large in astrophysical terms, but indicates that the presence of a magnetosphere may provide one further factor complicating the extraction of black hole parameters from the merger. The exploratory nature of this analysis, apparent in the approximations A1 and A2, limits the scope of broader conclusions, but relaxing both assumptions is technically feasible, as discussed in the next section.

%%%%%%%%%%%%%%%%%%%%%%%%%%%%%%%%%%%%%%%%%%%%%%%%%%%%%%%%%
\section{Concluding Remarks}\label{sec:conclusion}
%%%%%%%%%%%%%%%%%%%%%%%%%%%%%%%%%%%%%%%%%%%%%%%%%%%%%%%%%%
In this work, we carried out an exploratory analysis of the ringdown signatures of a Kerr black hole immersed in an asymptotically uniform magnetic field. This astrophysically relevant system was modeled using the Ernst-Wild geometry in the weak magnetic field limit, and the QNM spectrum was computed under some simplifying assumptions. In particular, we focused on slow rotation and scalar rather than tensor modes to simplify the analysis. These modes were then used to modify the usual Kerr template employed by LVK to obtain a modified linearized Ernst-Wild template. Several LVK events with high signal to noise for a ringdown-only analysis were then analyzed using the template to demonstrate how the presence of a magnetic field could potentially impact the extraction of the ringdown parameters.

Our results indicate consistency with the LVK analysis \cite{Abbott_2020} using the Kerr template, and the level of sensitivity to the magnetic field indicates that it will impact the extraction of parameters at current levels of sensitivity only if relatively large in units of the maximal field $B_M \sim 1/M$. However, our analysis has relied critically on the approximations A1 and A2, and it is natural to ask about the potential impact of relaxing these assumptions. It is apparent that the perturbative analysis at ${\cal O}(\tilde a)$ is a significant restriction given the current dataset, where the majority of events have relatively high spin, so that extending to ${\cal O}(\tilde a^2)$ would be valuable. This is technically possible, using recently developed approaches to coupled systems of perturbation equations \cite{Pani:2013pma}. However, we observe from Fig.~2 that up to $\tilde{a}\sim0.7$, the ${\cal O}(\tilde{a}^2)$ corrections to Kerr primarily impact Im$(\om)$ at the 10\% percent level. While most LVK events have relatively high spin, this may change with future observing runs (and eventually with the launch of space interferometers like LISA) potentially detecting events with smaller $\tilde a$, perhaps from black holes that have not merged before \cite{Gerosa:2017kvu}. Separately, based on the Kerr examples, switching from scalar to tensor modes may also induce non-negligible shifts in the QNM frequencies, and of course a fully quantitative analysis would require the physical tensor QNMs. Beyond the assumptions A1 and A2 that directly impact the ringdown template, there is also 
the problem that currently there is no information about the pre-merger system for producing the EW magnetized remnant, and specifically how the inspiral and merger may differ from the Kerr template. Further information about the evolution of the magnetosphere during the inspiral would also be needed for a more quantitative analysis, to gain insight into which QNMs are more likely to be excited in the remnant, and to better model the start time $t_0$ for the ringdown analysis.

%
%%%%%%%%%%%%%%%%%%%%%%%%
\begin{acknowledgments}
%%%%%%%%%%%%%%%%%%%%%%%%
We are particularly grateful to Gregorio Carullo for assistance with the implementation of \texttt{pyRing}, and to Michel Lefebvre for many valuable discussions regarding aspects of this work and for carefully reading the manuscript. We would also like to thank Emanuele Berti for illuminating discussions at the Testing Gravity 2023 conference, Frans Pretorius for providing additional insight at the Werner Israel Memorial Symposium, and Michalis Agathos for helpful comments at the poster session at the Kavli-Villum Summer School on Gravitational Waves. 
This work was supported in part by NSERC, Canada, and
has made use of data, software and/or web tools obtained from the Gravitational Wave Open Science Center (\url{https://www.gw-openscience.org/}), a service of the LIGO Laboratory, the LIGO Scientific Collaboration and the Virgo Collaboration. LIGO is funded by the U.S. National Science Foundation. Virgo is funded, through the European Gravitational Observatory (EGO), by the French Centre National de Recherche Scientifique (CNRS), the Italian Istituto Nazionale della Fisica Nucleare (INFN) and the Dutch Nikhef, with contributions by institutions from Belgium, Germany, Greece, Hungary, Ireland, Japan, Monaco, Poland, Portugal, Spain.
\textit{Software:} This work makes use of the following $\texttt{python}$ packages:
\texttt{pyRing, corner, cpnest, GWpy, lalsuite, matplotlib, numpy, scipy}, and \texttt{seaborn}
\cite{pyRing,cpnest, gwpy,lalsuite,matplotlib,numpy, scipy, seaborn}.
\end{acknowledgments}
\appendix

% change equation numbering
\counterwithin*{equation}{section}
\renewcommand\theequation{\thesection\arabic{equation}}
% changed equation numbering

%%%%%%%%%
\section{Ernst-Wild asymptotics at ${\cal O}(\tilde{B}^2)$}
%%%%%%%%%

In this Appendix, we provide further details of the large-$r$ asymptotic behaviour of Ernst-Wild perturbations when working perturbatively at a fixed order in $\tilde{B}$. For this purpose, it is helpful to first consider the pure Melvin spacetime, which determines the large-$r$ asymptotics of the Ernst-Wild geometry. The line element is given by setting  $M=\tilde a =0$ in Eq.~\ref{e:EW_metric},
\begin{equation}
    ds^2_M = \Lambda \left (  -dt^2 + dr^2+ r^2 d \theta^2 \right ) + \frac{\sin^2 \theta}{\Lambda} d\phi^2,
\end{equation}
where $\Lambda=\left(1-\frac{1}{4}B^2 r^2 \sin^2 \theta \right)^2$. In the absence of the black hole, it is more convenient to present the Melvin line element using cylindrical coordinates $\rho = r \sin \theta$ and $z= r \cos \theta$,
\begin{equation}
    ds^2_M = \Lambda (-dt^2 + d\rho^2 + dz^2) + \frac{\rho^2}{\Lambda}d \phi^2,
\end{equation}
with $\Lambda=\left (1-\frac{1}{4}B^2 \rho^2 \right )^2$.
This solution describes the back-reaction to a uniform magnetic field aligned along the $z$-direction, and reflects the non-asymptotically flat nature of the Ernst-Wild geometry at large radius. Following the approach of Section~\ref{sec:Perturbations_of_EW}, we can solve for massless scalar perturbations using Eq.~\ref{e:KG_scalar} with the following cylindrical ansatz,
\begin{equation}
    \label{e:KGCylindricalAnsatz}
    \Phi(t, \rho, z, \phi) = \frac{R(\rho)}{\sqrt{\rho}} e^{- i \omega t} e^{i m \phi} e^{i k z},
\end{equation}
and the Klein Gordon equation becomes \cite{Brito:2014super},
\begin{equation}
\begin{split}
    \label{e:MelvinODE}
    R''(\rho ) &- \frac{m^2 \left(B^2 \rho ^2+4\right)^4 R(\rho )}{256 \rho ^2} \\
    &-\frac{64 \left(4 k^2 \rho ^2-4 \rho ^2 \omega ^2-1\right)R(\rho )}{256 \rho ^2}=0.  
\end{split}
\end{equation}
The behaviour at infinity for the exact Melvin spacetime is given by
\begin{equation}
    R(\rho\rightarrow \infty) \rightarrow (B \rho)^{-3/2} e^{- (B \rho)^4 m /64},
\end{equation}
reflecting the confining asymptotics. However, in the regime $B \rho \ll 1$, and neglecting terms at $\mathcal O (B^4)$, 
the asymptotic boundary condition instead reduces to the form
\begin{equation}
    \label{e:BCMelvinPert2_wave}
    R(\rho\rightarrow \infty)|_{{\cal O}(B^2)} \rightarrow e^{i \rho  \sqrt{\omega ^2-k^2-B^2 m^2}},
\end{equation}
equivalent to an outgoing wave in Minkowski space, in contrast to the full Melvin asymptotics. Thus, provided we work perturbatively at ${\cal O}(B^2)$, perturbations are largely insensitive to the non-asymptotically flat geometry that sets in at large distance, and more precisely beyond the Melvin radius  $\rh_M = 2/B$. It follows that, after adding a black hole at the origin in this coordinate system, only the localized changes to the geometry near the horizon modify the mode solutions at ${\cal O}(B^2)$.

Returning to the full Ernst-Wild solution, the perturbative analysis above reflects the nontrivial transition in asymptotics in going from ${\cal O}(\tilde{B}^2)$ to ${\cal O}(\tilde{B}^4)$. Due to the long-range $1/r$ potential, the asymptotic boundary condition (\ref{e:Rellm(r)}) for perturbations at ${\cal O}(\tilde{B}^2)$ is equivalent to that for the Kerr spacetime, with the only correction appearing as a phase shift. Within the approximation (A2) outlined in Section~\ref{sec:Perturbations_of_EW}, 
we have neglected terms of ${\cal O}(\tilde{B}^4)$ that appear in the form $\tilde B^4 r^4$, thus the perturbation equation given by Eq.~\ref{Reqn} in the main text applies in the intermediate regime 
\be
 r  \ll r_M \sim 1/B.
 \ee  
We therefore have a notion of `infinity' close to $r/M =1/ \tilde B$, meaning that the model implicitly matches the Ernst-Wild metric far from the black hole to an asymptotically flat solution. However, when considering the perturbations that describe quasinormal modes this formal matching procedure is not necessary because the equation already exhibits the Kerr asymptotics due to the restriction to ${\cal O}(\tilde{B}^2)$.

\begin{figure}[t]
    \centering
    \includegraphics[width=\linewidth]{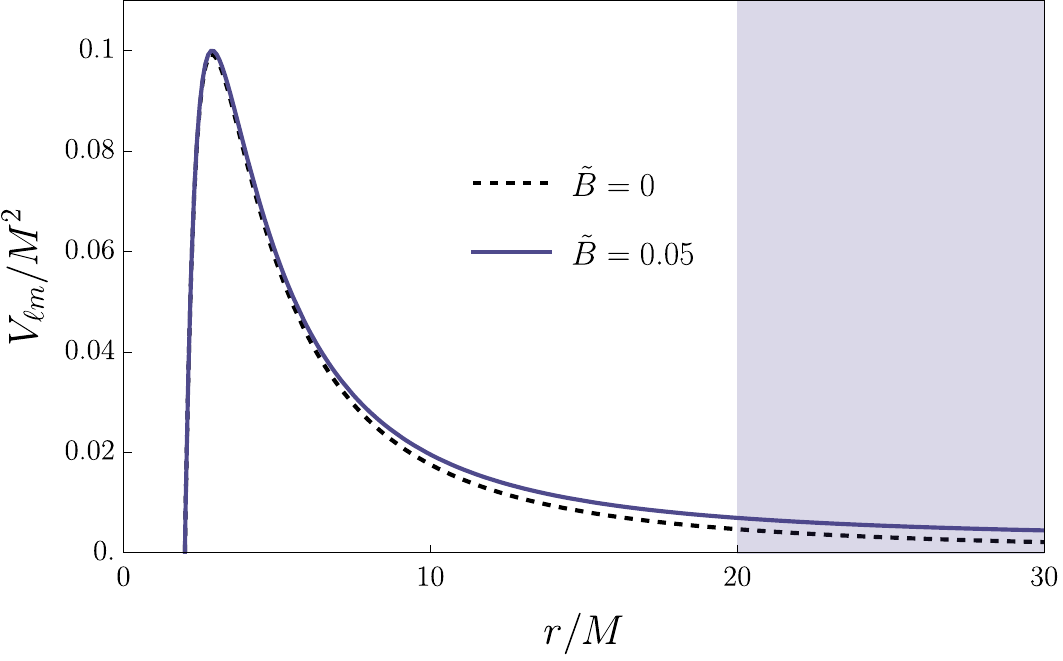}
    \caption{Radial effective potential for $(\ell, m) = (1,1)$ for the Schwarzschild metric ($\tilde B=0$) and the Ernst metric ($\tilde B = 0.05$). The shaded region is beyond the Melvin radius $\sim 1/B$, although the potential only receives a constant shift as $r\rightarrow \infty$ when restricted to ${\cal O}(\tilde{B}^2)$.}
    \label{fig:potential Schwarzschild  vs Ernst}
\end{figure}

Concretely, we can compare the perturbation potentials of the static (and non-asymptotically flat) Ernst geometry (with $\tilde{a}=0$) and the Schwarzschild geometry (with $\tilde{a}=\tilde{B}=0$) via Eq.~\ref{e:Veff_EW} in the main text,
\begin{equation}
    \label{e:Veff_Ernst}
    \begin{split}
    V_{\ell m }|_{{\cal O}(\tilde{B}^2)} = &-  \frac{(2M-r)\left(\ell (\ell +1) r+2 M\right)}{\,r^4}\\
    &-  \frac{(2M-r)\tilde B^2 m^2 r^3}{M^2 r^4}.
    \end{split}
\end{equation}
The effective potential for the Schwarzschild black hole on the first line is only corrected by a constant at large radius at ${\cal O}(\tilde{B}^2)$. The shift is exhibited in Fig.~\ref{fig:potential Schwarzschild vs Ernst}, with the shaded region showing the domain beyond the Melvin radius where the asymptotics would be modified more substantially at ${\cal O}(\tilde{B}^4)$. The restriction to ${\cal O}(\tilde{B}^2)$ thus restores the expectation within the Kerr spacetime that the QNM spectrum is primarily affected by the behaviour of the potential near the horizon and photon sphere of the black hole \cite{Konoplya:2006ar,Konoplya:2006rv,Konoplya:2007yy,Konoplya:2008hj,Brito:2014super}.

\bibliography{refs.bib}

\end{document}